\documentclass[11pt]{article}
\usepackage{amsmath}
\usepackage{graphicx,psfrag,epsf,booktabs,setspace}
\usepackage{enumerate}
\usepackage{natbib}
\usepackage{url}
\usepackage[flushleft]{threeparttable}
\usepackage{ amssymb }
\usepackage{ upgreek }
\usepackage{float} 
\usepackage[title]{appendix}

\newtheorem{myDef}{Definition}

\usepackage{multirow}

\begin{document}

\def\spacingset#1{\renewcommand{\baselinestretch}%
{#1}\small\normalsize} \spacingset{1}

  \title{\bf Measures of Model Risk in Continuous-time Finance Models}
\author{
Emese Lazar\footnote{ICMA Centre, Henley Business School, University of Reading, Whiteknights, Reading, RG6 6BA, United Kingdom; e.lazar@icmacentre.ac.uk.}, Shuyuan Qi\footnote{
ICMA Centre, Henley Business School, University of Reading, Whiteknights, Reading, RG6 6BA, United Kingdom; s.qi@pgr.reading.ac.uk.}, Radu Tunaru\footnote{Corresponding author, University of Sussex Business School, University of Sussex, Brighton, BN1 9SL, UK; R.Tunaru@sussex.ac.uk.}\\
}\singlespacing
  \maketitle

\vspace{-.2in}
\begin{abstract}
     Measuring model risk is required by regulators on financial and insurance markets. We separate model risk into parameter estimation risk and model specification risk, and we propose expected shortfall type model risk measures applied to L\'{e}vy jump models and affine jump-diffusion models. We investigate the impact of parameter estimation risk and model specification risk on the models' ability to capture the joint dynamics of stock and option prices. We estimate the parameters using Markov chain Monte Carlo techniques, under the risk-neutral probability measure and the real-world probability measure jointly. We find strong evidence supporting modeling of price jumps.
\end{abstract}

\medskip
\noindent%

\thispagestyle{empty}
\vfill
\clearpage 

\spacingset{1.5} 
\setcounter{footnote}{0}
\renewcommand{\thefootnote}{\arabic{footnote}}

\section{Introduction}
    Model risk is currently considered one of the most overlooked risks faced by financial firms. The \cite{basle2009revisions}, \cite{reserve2011supervisory} and \cite{eba2012mr} require banks to measure and report model risk as with any other type of risk. The sources of model risk are parameter estimation risk (PER) and model specification risk (MSR). The specification of models includes identifying and modeling decisive factors that can jointly describe the dynamics of an economic asset. PER denotes the risk of inaccurate estimation of parameters for a given model. The  PER or MSR are the two components of the total model risk (TMR).\footnote{In this paper, we use the following abbreviations: PER = parameter estimation risk; MSR = model specification risk; TMR= total model risk; ES = expected shortfall; VaR = value-at-risk; AJD = affine jump-diffusion; MCMC = Markov chain Monte Carlo; SV = stochastic volatility; SVJ = stochastic volatility with Merton jumps in returns; SVCJ = stochastic volatility with contemporaneous jumps in returns and volatility; SVVG = stochastic volatility with variance-gamma jumps in returns; SVLS = stochastic volatility with log-stable jumps in returns; MJD = Merton jump-diffusion; SND = standard normal distribution; ATM = at-the-money; KS = Kolmogorov-Smirnov; DIC = deviance information criterion; log-BF = log values of the Bayes factors; PE = pricing error; APE = absolute pricing error; DM = Diebold and Mariano; CW = Clark and West.} 

    The majority of studies use point-wise estimation methods and consider model risk as model mispricing, thus ignoring parameter estimation risk.
    \cite{kerkhof2010model} is one notable exception where model risk is separated into PER and MSR, but their approach relies on gaussian returns and the asymptotic distribution of a function of the parameters.
    PER can be captured via Bayesian estimation methods. \cite{Jacquier2000} study the PER of the Black and Scholes model. Furthermore, \cite{jacquier2002bayesian} apply Bayesian estimators for stochastic volatility models and find that the Bayesian approach produces more robust results by comparison with the moments and likelihood estimators. There are many other advantages of using Markov chain Monte Carlo (MCMC) techniques in extracting inference on continuous-time models in finance that have been highlighted in a series of works by \cite{Eraker2001}, \cite{PolsonStroud}, \cite{jacquier2007mcmc}, \cite{johannes2010mcmc} and \cite{yu2011mcmc}.
 
    In this paper, we propose an expected shortfall (ES) based method to measure model risk. This ES-type model risk measurement is potentially superior in capturing model risk because it is able to capture model tail risk. We provide an applicable framework to separate and measure PER and MSR for a very competitive class of option pricing models. In addition, we disentangle the model risk for buyers and sellers and we highlight that the two parties in options contracts are exposed to model risk asymmetrically.
    
    We apply our new methodology to several different option pricing models with respect to their modeling ability to explain S\&P 500 spot prices and option prices. The candidate models that we investigate have different specifications: the constant volatility model with Merton jumps (\citeauthor{merton1976option}, \citeyear{merton1976option}); the pure stochastic volatility controlled by a mean-reversion process; as well as the stochastic volatility with affine jump-diffusion (AJD) or L\'{e}vy jumps. Consistent with the findings of \cite{yu2011mcmc}, we find that the log-stable jumps model with stochastic volatility process has the smallest TMR. Moreover, the TMR of AJD models is mainly attributed to PER, and their MSR is less than that of L\'{e}vy jump models. The infinite-activity L\'{e}vy jumps capture many small jumps in the index returns that cannot be captured by AJD models. Therefore, L\'{e}vy jump models may face less MSR during turbulent periods when small jumps are frequent; however, this also restricts their ability in capturing both physical and risk-neutral dynamics when the market is calm and there are far fewer jumps. In our paper, we find that the seller of vanilla European call options is exposed to a higher model risk than the buyer of those options when the market is volatile, which is in line with the conclusions of \cite{green1999market}.

    The remainder of the paper is organized as follows. In Section \ref{mrss}, we introduce the model risk measurement framework; then we revisit all models that are investigated in the paper in Section \ref{Pricingmodels}. Section \ref{nm} provides the details of the numerical methods applied in this work. An  empirical analysis is presented in Section \ref{EA} whilst in Section \ref{ra} we analyze the forecasting of model risk. The last section summarizes our conclusions.

\section {Model Risk}\label{mrss}

     Most literature in the area of model risk of continuous-time models concern measuring TMR.
    \cite{Routledge2009} distinguish between market risk and model risk; they measure market risk by risk-neutral pricing and measure model risk by using the worst-case approach.
   Further examples are the coherent and convex risk measures in \cite{cont2006model}. 
   Another innovative  work on model risk is \cite{lindstrom2008sequential}, where parameter uncertainty is taken into account with a revised risk-neutral valuation formula. Furthermore, \cite{detering2016model} measure the model risk of option pricing models using a hedging portfolio argument.
    \cite{coqueret2016investigation} show that model selection can lead to a significant effect on the final results since the discrepancy in prices leads to model risk. Moreover, \cite{chen2011generalized} emphasize the importance of model specification; they develop an omnibus specification test for continuous-time models and study the AJD and L\'{e}vy processes specifically.
    
Using a Bayesian approach, one can obtain the estimated posterior distribution of an asset price, denoted as $\widetilde{F_t}( \mathcal{H};\mathcal{M}(\Theta), \mathcal{D}, \mathcal{K})$. In this case  $\mathcal{H}$ is the option (call or put) conditional on model $\mathcal{M}$ with parameter vector $\Theta$ at time $t$, given an observed dataset $\mathcal{D}$ covering the historical series of the option prices and underlying asset observations. For clarity we also  insist on the notational $\mathcal{K}$ for different computational methodologies (including estimation, calibration and pricing). The posterior distribution of option prices is produced by the uncertainty in the value of parameters $\Theta$ weighted by the combination of prior assumptions on $\Theta$ and the likelihood coming out of historical data.

   We refer to the risk stemming from model specification as MSR, while the risk related to parameter estimation is referred to as PER. For a set of observed data $\mathcal{D}$ and a set of consistent methodologies $\mathcal{K}$, we define PER, MSR and TMR as follows.

    \begin{myDef}\label{perd}
         For option $\mathcal{H}$ and model $\mathcal{M}$ with the vector of parameters $\Theta$, the parameter estimation risk refers to the uncertainty in the values of parameters $\Theta$ obtained via the estimation process $\mathcal{K}$ given dataset $\mathcal{D}$.
    \end{myDef}
    
    A model can be misspecified for many reasons, for example, because it ignores significant factors and fails to capture the features of the market fully; and/or its assumptions are unrealistic (e.g., the constant-volatility assumption in the Black-Scholes options pricing model). The MSR measures the risk due to the inherent weakness of the model itself to get the correct results.
    \begin{myDef}\label{msrd}
         For option $\mathcal{H}$ and model $\mathcal{M}$ with the vector of parameters $\Theta$, the model specification risk of model $\mathcal{M}$ refers to the risk that, based on dataset $\mathcal{D}$ and methodologies $\mathcal{K}$, the model
         is unable to produce the features of $\mathcal{H}$.
    \end{myDef}

The TMR is defined as the sum of the two components. Using the information in the option markets, \cite{jarrow2015specification} point out that it is challenging to separate model specification risk from estimation risk. Given that our focus is on the model risk of option pricing models, under the physical and risk-neutral probability measures defining the financial markets, we first introduce three properties that a valid model risk measure for option pricing models should have:
 (1) \emph{time variability}: model risk is time-varying; 
 (2) \emph{symmetry of MSR}: the MSR is the same for both long and short positions; 
 (3) \emph{asymmetry of PER}: the PER can be different for long and short positions.\footnote{Theoretically, a short position for an option leads to a larger PER. As the option price ranges from 0 to infinity, the left side is bounded while the right side is open, which might lead to a broader right tail of the posterior densities of the estimated price. Options are affected by the asymmetry between buying and writing, in that the option buyer has liability limited to the amount invested, but the option writer is exposed to the risk of losses that can greatly exceed the initial premium received (\citeauthor{green1999market}, \citeyear{green1999market}).}

   Our main objective is to measure PER and MSR, respectively, and then to compare the model risk across different pricing models. \cite{johannes2010mcmc} state that the marginal posterior distribution through the Bayesian estimation characterizes the sample information regarding the objective and risk-neutral parameters and quantifies the parameter estimation risk. \cite{chung2013explaining} take the posterior distribution through the Bayesian approach as relevant for PER.

    Let $\widehat{F}$ represent an estimated price of the target option,\footnote{$\widetilde{F}$ denotes the estimated price distribution and $\widehat{F}$ is a point estimate.} the model price adjusted posterior distribution $\widetilde{\Lambda}$ for long ($L$) and short ($S$) positions is defined as:
    \begin{equation}\label{pere}
    \begin{split}
     &\widetilde{\Lambda_{t}}( \mathcal{H}, L;\mathcal{M}(\Theta), \mathcal{D}, \mathcal{K}) =\widetilde{F_t}( \mathcal{H};\mathcal{M}(\Theta), \mathcal{D}, \mathcal{K})  - \widehat{F_t}( \mathcal{H};\mathcal{M}(\Theta), \mathcal{D}, \mathcal{K}) , \\
     &\widetilde{\Lambda_{t}} ( \mathcal{H},S;\mathcal{M}(\Theta), \mathcal{D}, \mathcal{K})  = \widehat{F_t}( \mathcal{H};\mathcal{M}(\Theta), \mathcal{D}, \mathcal{K})  - \widetilde{F_t}( \mathcal{H};\mathcal{M}(\Theta), \mathcal{D}, \mathcal{K}) .
    \end{split}
    \end{equation}

For example, if we consider the expected value of the posterior price distribution with respect to $\Theta$ as the estimated price, then the above formulae can be expressed as:
        \begin{equation}\label{pere——2}
    \begin{split}
     &\widetilde{\Lambda_{t}}( \mathcal{H},L;\mathcal{M}(\Theta), \mathcal{D}, \mathcal{K})  =\widetilde{F_t}( \mathcal{H};\mathcal{M}(\Theta), \mathcal{D}, \mathcal{K})  - {\rm E}_\Theta \left[\widetilde{F_t}( \mathcal{H};\mathcal{M}(\Theta), \mathcal{D}, \mathcal{K}\right] , \\
     &\widetilde{\Lambda_{t}} ( \mathcal{H},S;\mathcal{M}(\Theta), \mathcal{D}, \mathcal{K})  = {\rm E}_\Theta \left[\widetilde{F_t}( \mathcal{H};\mathcal{M}(\Theta), \mathcal{D}, \mathcal{K}) \right] - \widetilde{F_t}( \mathcal{H};\mathcal{M}(\Theta), \mathcal{D}, \mathcal{K}).
    \end{split}
    \end{equation}

     Let $VaR_{\eta,\,t}^{PER}( \mathcal{H}, \hbar;\mathcal{M}(\Theta), \mathcal{D}, \mathcal{K}) $ denote the value-at-risk (VaR) at a critical level\footnote{We consider $\eta$ to be 5\% in this paper.} $\eta \in (0,1)$, which is computed as the absolute value of the $\eta$ quantile of the adjusted posterior distribution $\widetilde{\Lambda_{t}}( \mathcal{H},\hbar;\mathcal{M}(\Theta), \mathcal{D}, \mathcal{K})$ computed in (\ref{pere——2}), where $\hbar=L$ for a long position and $\hbar=S$ for a short position. The ES-type model risk measure for PER, at level $\eta$, for option $\mathcal{H}$, given a model $\mathcal{M}$ with parameter vector $\Theta$, dataset $\mathcal{D}$, methodology $\mathcal{K}$, and a long(short) position $\hbar$ is defined as:
    \begin{equation}\label{perre1}
    \rho^{PER}_{\eta, \,t} ( \mathcal{H},\hbar;\mathcal{M}(\Theta), \mathcal{D}, \mathcal{K})  =  \frac{1}{\eta} \int_{0}^{\eta} VaR_{x, \,t}^{PER}( \mathcal{H},\hbar;\mathcal{M}(\Theta), \mathcal{D}, \mathcal{K}) dx.
    \end{equation}

    The market price of the option at time $t$ is $C_t(\mathcal{H})$. We can further compute the distribution of profit and loss from pricing with the model for both long and short positions by using the linear functions below:
    \begin{equation}\label{tomr}
    \begin{split}
    &\Lambda_{t}( \mathcal{H},L;\mathcal{M}(\Theta), \mathcal{D}, \mathcal{K}) =\widetilde{F_t}( \mathcal{H};\mathcal{M}(\Theta), \mathcal{D}, \mathcal{K})- C_t(\mathcal{H}), \\
    &\Lambda_{t}( \mathcal{H},S;\mathcal{M}(\Theta), \mathcal{D}, \mathcal{K}) = C_t(\mathcal{H}) - \widetilde{F_t}( \mathcal{H};\mathcal{M}(\Theta), \mathcal{D}, \mathcal{K}).
    \end{split}
    \end{equation}

    The VaR of the profit and loss distribution at level $\eta$ of a long(short) position, denoted by $VaR_{\eta, \,t}( \mathcal{H}, \hbar;\mathcal{M}(\Theta), \mathcal{D}, \mathcal{K})$, is computed as the absolute value of the $\eta$ quantile of $\Lambda_{t}( \mathcal{H},\hbar; \mathcal{M}(\Theta), \mathcal{D}, \mathcal{K})$. The ES-type model risk measure of the TMR for option $\mathcal{H}$, given a model $\mathcal{M}$ with parameter vector $\Theta$, dataset $\mathcal{D}$, and methodology $\mathcal{K}$ is defined as:
    \begin{equation}\label{tomr3}
    \rho^{TMR}_{\eta, \,t} ( \mathcal{H};\mathcal{M}(\Theta), \mathcal{D}, \mathcal{K})= \frac{ \int_{0}^{\eta} VaR_{x, \,t} (\mathcal{H}, L;\mathcal{M}(\Theta), \mathcal{D}, \mathcal{K}) dx + \int_{0}^{\eta} VaR_{x, \,t}( \mathcal{H},S;\mathcal{M}(\Theta), \mathcal{D}, \mathcal{K})dx }{2\eta}.
    \end{equation}

    The PER of option $\mathcal{H}$, given a model $\mathcal{M}$ with parameter vector $\Theta$, dataset $\mathcal{D}$, and methodology $\mathcal{K}$ is defined as the average of the PER values for long and short positions.
    \begin{equation}\label{perre}
    \rho^{PER}_{\eta, \,t}( \mathcal{H};\mathcal{M}(\Theta), \mathcal{D}, \mathcal{K})=\frac{\rho^{PER}_{\eta, \,t} ( \mathcal{H},L;\mathcal{M}(\Theta), \mathcal{D}, \mathcal{K})+\rho^{PER}_{\eta, \,t} ( \mathcal{H},S;\mathcal{M}(\Theta), \mathcal{D}, \mathcal{K})}{2}.
    \end{equation}

    Then, the MSR of option $\mathcal{H}$, given a model $\mathcal{M}$ with parameter vector $\Theta$, dataset $\mathcal{D}$, and methodology $\mathcal{K}$ is measured as the difference between TMR and PER:
    \begin{equation}\label{msrdcrother}
    \rho^{MSR}_{\eta, \,t} ( \mathcal{H};\mathcal{M}(\Theta), \mathcal{D}, \mathcal{K}) = \rho^{TMR}_{\eta, \,t}( \mathcal{H};\mathcal{M}(\Theta), \mathcal{D}, \mathcal{K}) - \rho^{PER}_{\eta, \,t}( \mathcal{H};\mathcal{M}(\Theta), \mathcal{D}, \mathcal{K}).
    \end{equation}

    Because of the symmetry of MSR and asymmetry of PER, the TMR for $\hbar$ is the sum of PER for $\hbar$ and MSR, where $\hbar=L$ or $\hbar=S$:
     \begin{equation}\label{tmrLS}
    \rho^{TMR}_{\eta, \,t} ( \mathcal{H},\hbar;\mathcal{M}(\Theta), \mathcal{D}, \mathcal{K})= \rho^{PER}_{\eta, \,t} ( \mathcal{H},\hbar;\mathcal{M}(\Theta), \mathcal{D}, \mathcal{K}) +  \rho^{MSR}_{\eta, \,t} ( \mathcal{H};\mathcal{M}(\Theta), \mathcal{D}, \mathcal{K}).
     \end{equation}

     Based on (\ref{perre}), (\ref{msrdcrother}) and (\ref{tmrLS}), the following holds:
    \begin{equation}\label{relation}
     \rho^{TMR}_{\eta, \,t} ( \mathcal{H};\mathcal{M}(\Theta), \mathcal{D}, \mathcal{K}) = \frac{\rho^{TMR}_{\eta, \,t} ( \mathcal{H},L;\mathcal{M}(\Theta), \mathcal{D}, \mathcal{K}) + \rho^{TMR}_{\eta, \,t} ( \mathcal{H},S;\mathcal{M}(\Theta), \mathcal{D}, \mathcal{K})}{2}.
    \end{equation}

   It is important to note that
   $\rho^{TMR}_{\eta, \,t} ( \mathcal{H},\hbar;\mathcal{M}(\Theta), \mathcal{D}, \mathcal{K}) \neq \frac{1}{\eta} \int_{0}^{\eta} VaR_{x, \,t}( \mathcal{H},\hbar;\mathcal{M}(\Theta), \mathcal{D}, \mathcal{K}) dx$,
   unless $C_t(\mathcal{H}) = {\rm E}_\Theta \left[\widetilde{F_t}( \mathcal{H};\mathcal{M}(\Theta), \mathcal{D}, \mathcal{K}) \right]$, and that in this case $\rho^{MSR}_{\eta, \,t} ( \mathcal{H};\mathcal{M}(\Theta), \mathcal{D}, \mathcal{K}) = 0$. Moreover, if the model price estimate of $\mathcal{H}$ is close enough to the market price, PER is the main source of model risk. However, MSR is the primary source of model risk if the market price of $\mathcal{H}$ is far from the model price estimate. This is a desirable property of our proposed model risk measurement.

\section{Models}\label{Pricingmodels}
This section describes the set of models that will be compared: the stochastic volatility (SV) model, the stochastic volatility model with Merton jumps in returns (SVJ), the stochastic volatility model with contemporaneous jumps in returns and volatility (SVCJ), the stochastic volatility model with variance-gamma jumps in returns (SVVG) and the stochastic volatility model with log-stable jumps in returns (SVLS). While the model specification is expressed under the physical measure $\mathbb{P}$,  it is the return dynamics under risk-neutral measure $\mathbb{Q}$ which is required for option pricing. The change of measure between $\mathbb{P}$ and $\mathbb{Q}$ for these models is also discussed in this section.

\subsection{Stochastic Volatility Models}\label{model_return}

Let $Y_t=\ln(S_t)$ denote the logarithm of the asset price. The dynamics, for all models, of the continuously compounded return on the asset price under the real-world measure $\mathbb{P}$ is given by:
    \begin{equation}\label{SV_models}
    \begin{split}
        &dY_t = \mu dt + \sqrt{V_{t}} dW_t^Y(\mathbb{P}) + dJ_t^Y(\mathbb{P}),\\
        &dV_t = \kappa (\theta - V_t)dt + \sigma_V \sqrt{V_t} dW_t^V(\mathbb{P}) + dJ_t^V(\mathbb{P}),
    \end{split}
    \end{equation}
where $W_t^Y(\mathbb{P})$ and $W_t^V(\mathbb{P})$  are standard Brownian motions under $\mathbb{P}$ with $dW_t^Y(\mathbb{P}) dW_t^V(\mathbb{P}) = \rho dt$, the correlation $\rho$ provides the ability to capture the skewness of the returns' distribution. A negative $\rho$ captures the leverage effect; $\mu$ measures the mean return; $V_t$ is the instantaneous variance of returns at time $t$; $\kappa$ represents the speed of mean reversion; $\theta$ denotes the long-run mean of the variance process; and $\sigma_V$ is the volatility of volatility.

The SVJ and SVCJ are AJD models (\citeauthor{duffie2000transform}, \citeyear{duffie2000transform}) capturing continuous movements of assets with affine diffusions and  large discontinuous jumps in asset returns with a Poisson process. The jump processes in AJD models are defined as $dJ_t^Y(\mathbb{P}) = \upxi^Y dN_t^Y$ and $dJ_t^V(\mathbb{P}) = \upxi^V dN_t^V$ where $\{N_t^Y\}_{t \geq 0} $ and $\{N_t^V\}_{t \geq 0}$ are Poisson processes as in \cite{duffie2000transform}. The SVCJ model contains simultaneous correlated jumps, where\footnote{This is a standard setting; large movements in equity returns and large shifts in the variance are likely to occur at the same time (\citeauthor{bardgett2019inferring}, \citeyear{bardgett2019inferring}).} $N_t^Y = N_t^V = N_t$, in both the return and volatility processes with a constant intensity\footnote{\cite{bates2000post} finds that the model with state-dependent intensities is significantly misspecified whilst \cite{andersen2002empirical} state that there is no evidence to support the time-varying intensity.} $\lambda$; the jump size in the variance process follows an exponential distribution, $\upxi^V \sim \mathbb{EXP} (\mu_V)$, and the jump size in the asset log-prices is conditionally normally distributed with $\upxi^Y| \upxi^V\sim \mathbb{N}(\mu_J + \rho_J \upxi^V ,\sigma_J^2)$.\footnote{$\mathbb{EXP}$ denotes the exponential distribution and $\mathbb{N}$ denotes the normal distribution.} Thus, the mean of $\upxi^Y$ is $\mu_J + \rho_J \mu_V$ and its variance is $\sigma_J^2 + \rho_J^2 \mu_V^2$. Moreover, $\upxi^Y$ is correlated with $\upxi^V$ by $\rho_J \mu_V / \sqrt{\sigma_J^2 + \rho_J^2 \mu_V^2}$, the long-run mean-variance of the SVCJ model is $\theta + \mu_V \lambda /\kappa$ due to the jump component in the variance process. As explained in \cite{ball1985jumps}, the return distribution is an infinite mixture of normal distributions under the jump-diffusion models, which leads to an unbounded likelihood function. To circumvent this issue, we assume that only one jump occurs per trading day. For SVJ, $J_t^V(\mathbb{P})=0$, and the process of the jump $J_t^Y(\mathbb{P})$ has the same specification as SVCJ. For SV, $J_t^Y(\mathbb{P})=J_t^V(\mathbb{P})=0$.

The AJD model is constructed based on Brownian motions and compound Poisson processes, which are just special cases of L\'{e}vy processes. AJD models only allow finite-activity jump processes, while the L\'{e}vy processes are more flexible, allowing them to achieve infinite jump arrival rates. In our empirical analysis, we consider  two L\'{e}vy jump models, namely the SVVG model in \cite{madan1998variance} and the SVLS model in \cite{carr2003finite}. 

The SVVG model is a kind of infinite-activity but finite-variation jump model. The variance gamma process can be described  as:
\begin{equation}\label{VG}
   X^{VG}_{t}(\sigma, \gamma, \nu) = \gamma G_t^{\nu} + \sigma W_{G_t^\nu},
\end{equation}
where $\{X^{VG}\}$ is an arithmetic Brownian motion with drift $\gamma$ and volatility $\sigma$; $\{G_t^{\nu}\}_{t\geq 0}$ denotes the gamma process with unit mean rate and variance rate $\nu$; and $\{W_t\}_{t \geq 0}$ is a standard Brownian motion, which is independent of $G_t^{\nu}$. Setting $J_t^Y(\mathbb{P})=X^{VG}_{t}(\sigma, \gamma, \nu)$ and $J_t^V(\mathbb{P})=0$ reduces (\ref{SV_models}) to SVVG. 

The SVLS model is an example of infinite-activity and infinite-variation jump model. The log-stable process follows an $\alpha$-stable distribution ($S_\alpha$):
\begin{equation}\label{LS}
  X^{LS}_{t}(\alpha, \sigma) - X^{LS}_{s}(\alpha, \sigma) \sim S_\alpha (\beta,  \delta(t-s)^\frac{1}{\alpha},\gamma),\,\, t>s,
\end{equation}
where $\alpha \in (1, 2]$ is the tail index of the $\alpha$-stable distribution  which determines the shape of the stable distribution; $\beta \in [-1, 1]$ is the skew parameter determining the skewness of the distribution; $\delta \geq -1$ is the scale parameter and $\gamma \in \mathbb{R}$ is the location parameter. We follow \cite{carr2003finite} and set $\beta = 0$, $\delta = \sigma$ and $\gamma=0$ and then (\ref{SV_models}) reduces to SVLS, if $J_t^Y(\mathbb{P})=X^{LS}_{t}(\alpha, \sigma)$ and $J_t^V(\mathbb{P})=0$.

\subsection{Change of Measure and Option Pricing for Stochastic Volatility Models}\label{model_op}

For Brownian motions, \cite{pan2002jump} proposes a standard practice for the change of measure:
\begin{equation}\label{Change_measure}
    \begin{split}
   &\gamma_t^{Y} = \eta_s\sqrt{V_t},\\
   &\gamma_t^{V} = -\frac{1}{\sqrt{1-\rho^2}} \left(\rho \eta_s + \frac{\eta_v}{\sigma_V}\right)\sqrt{V_t},
    \end{split}
\end{equation}
where $\gamma_t^{Y}$ and $\gamma_t^{V}$ represent the market prices of risk of Brownian shocks to returns and variance, respectively. The Brownian motions under $\mathbb{Q}$ in the return and variance processes are:
\begin{equation}\label{BM_Q}
    \begin{split}
   &dW_t^Y(\mathbb{Q}) = dW_t^Y(\mathbb{P}) + \gamma_t^{Y}dt,\\
   &dW_t^V(\mathbb{Q}) = dW_t^V(\mathbb{P}) + \gamma_t^{V}dt.
    \end{split}
\end{equation}

 For the variance process, following \cite{pan2002jump}, \cite{bates2000post} and \cite{broadie2007model}, we apply the following theoretical restrictions in the change of measure such that both physical and risk-neutral probability densities are from the same family:
$\kappa^{\mathbb{P}} \theta^{\mathbb{P}}  = \kappa^{\mathbb{Q}} \theta^{\mathbb{Q}}$; $\rho^{\mathbb{P}} = \rho^{\mathbb{Q}}$; and $\sigma_V^{\mathbb{P}} = \sigma_V^{\mathbb{Q}}$.\footnote{$\kappa^{\mathbb{Q}} = \kappa^{\mathbb{P}} - \eta_v$ and $\theta^{\mathbb{Q}} = \frac{\kappa^{\mathbb{P}} \theta^{\mathbb{P}}}{\kappa^{\mathbb{Q}}}$, we use $\kappa$ and $\theta$ to represent $\kappa^{\mathbb{P}}$ and $\theta^{\mathbb{P}}$ in this paper. Moreover, for simplicity, we use $\rho$ to denote $\rho^{\mathbb{P}}$ and $\rho^{\mathbb{Q}}$, and use $\sigma_V$ to represent $\sigma_V^{\mathbb{P}}$ and $\sigma_V^{\mathbb{Q}}$.}

Moreover, for jump processes, the following restrictions are imposed: in SVJ, $(\lambda, \mu_J, \sigma_J)$ are able to change between $\mathbb{P}$ and $\mathbb{Q}$; in SVCJ, $(\lambda, \mu_J, \sigma_J, \rho_J, \mu_V)$ are able to change between $\mathbb{P}$ and $\mathbb{Q}$; in SVVG, $\gamma$ and $\sigma$ are able to change between $\mathbb{P}$ and $\mathbb{Q}$, while $\nu$ remains unchanged under $\mathbb{P}$ and $\mathbb{Q}$; in SVLS, no parameters of the log-stable process, $(\alpha, \beta, \sigma, \gamma)$, are allowed to change between $\mathbb{P}$ and $\mathbb{Q}$.

In AJD models, all parameters in the jump processes can be different under the physical and risk-neutral measures. However, this leads to difficulty in econometric identification, as shown in \cite{pan2002jump} and \cite{eraker2004stock}. To bypass this identification issue, they only enable $\mu_J$ to change between measures. We adopt the same methodology here. Finally,  the jump parameters under both measures for SVJ, SVCJ, SVVG and SVLS are $(\lambda, \mu_J^\mathbb{P}, \sigma_J, \mu_J^{\mathbb{Q}})$, $(\lambda, \mu_J^\mathbb{P}, \sigma_J, \mu_J^{\mathbb{Q}}, \mu_V, \rho_J)$, $(\nu, \gamma^\mathbb{P}, \sigma^\mathbb{P}, \gamma^{\mathbb{Q}}, \sigma^{\mathbb{Q}})$ and $(\alpha, \sigma)$, respectively.

Under the framework described above, the Radon-Nikodym derivatives of theses processes can be expressed as:\footnote{$U_t$ in (\ref{RN_deri}) is defined in the second part of \cite{sato1999levy}'s theorem. As jump processes are restricted to follow the same processes between measures, the $U_t$'s of models with different jump specifications are considered. In this case, models also differ in terms of Radon-Nikodym derivatives. Additionally, in this paper we use both $e$ and $\exp$ to denote the exponential function.}
\begin{equation}\label{RN_deri}
    \begin{split}
   \left. \frac{d\mathbb{Q}}{d\mathbb{P}}\middle| \right._t = &\exp \left\{ -\int_{0}^{t}\gamma_s^YdW_s^Y(\mathbb{P}) - \int_{0}^{t}\gamma_s^VdW_s^V(\mathbb{P})\right.\\
   &\left.- \frac{1}{2} \left[\int_{0}^{t} (\gamma_s^Y)^2 ds + \int_{0}^{t} (\gamma_s^V)^2 ds\right] \right\} \exp(U_t).
    \end{split}
\end{equation}

Then the return dynamics under the risk-neutral measure are expressed as:
\begin{equation}\label{R_N_E}
    \begin{split}
    &dY_t = \left(r_t - \frac{1}{2}V_t + \Phi_J(-i)\right) dt + \sqrt{V_t}dW_t^Y(\mathbb{Q}) + dJ_t^Y(\mathbb{Q}),\\
    &dV_t = \left[\kappa (\theta - V_t) + \eta_v V_t \right]dt + \sigma_V \sqrt{V_t} dW_t^V(\mathbb{Q}) + dJ_t^V(\mathbb{Q}),
    \end{split}
\end{equation}
where $r_t$ is the risk-free rate, $\Phi_J(-i)$ is the jump component, and the expressions of $\Phi_J(\centerdot)$ for different models in this study are documented in Appendix \ref{appA}. Naturally, the drift term of the return process under $\mathbb{P}$ can be derived as $\mu = r_t - \frac{1}{2}V_t + \Phi_J(-i) + \eta_s V_t$.

When the interest rate is constant, the option price can be deduced from the closed-form solution to the characteristic function of the log stock price under $\mathbb{Q}$: 
\begin{equation}\label{cha}
    \phi(t, u)= \exp\left[{iuY_0+iu(r+\Phi_J(-i))t}\right] \exp\left({-t\Phi_J(u)}\right) \exp\left({-b(t, u)V_0 - c(t, u)}\right),
\end{equation}
where $\kappa^M(u) = \kappa - \eta_v - iu\sigma_V \rho$; $\delta(u) = \sqrt{(\kappa ^M(u))^2 + (iu + u^2)\sigma_V^2}$; $Y_0 = \ln(S_0)$ denotes log-spot price; $V_0$ represents the initial variance;
$b(t, u) = \frac{(iu+u^2)(1-e^{-\delta(u) t})}{(\delta(u) + \kappa^M(u)) + (\delta(u) - \kappa^M(u))e^{(-\delta(u) t)}}$; and $c(t, u) = \frac{\kappa \theta}{\sigma_V^2} \left[ 2 \ln \frac{2 \delta(u) -(\delta(u) -\kappa^M(u) )(1-e^{-\delta(u) t})}{2 \delta(u)} + (\delta(u) - \kappa^M(u))t \right]$. 
     
Taking $\tau$ as the time to expiration, one can then price a European call option with strike $K$, using the formula below (\citeauthor{yu2011mcmc}, \citeyear{yu2011mcmc}):
\begin{equation}\label{pricing}
    F(Y_0, V_0, \tau, K) = {\rm E}_0^{\mathbb{Q}}[e^{-r \tau}(S_\tau - K)^{+}] = \frac{e^{-r \tau}}{\pi} \times {\rm Re}\left( \int_{0}^{\infty} e^{-ix \ln(K)} \frac{\phi (\tau, x-i)}{-x^2 + ix} dx \right).
\end{equation}

\section{Estimation Methods} \label{nm}

      The models analyzed in this paper are quite complex so parameter estimation may not be straightforward. One problem is that the latent variables, such as stochastic volatility and size and arrival rates of jump processes, are difficult to track.  L\'{e}vy processes themselves are complex and many of them  do not lead to  closed-form option pricing formulae. \cite{li2006bayesian} extend the application of the MCMC method to L\'{e}vy processes under the real-world probability measure ($\mathbb{P}$) for  spot prices. Then, \cite{yu2011mcmc} further apply the MCMC method to L\'{e}vy processes under both $\mathbb{P}$ and $\mathbb{Q}$ probability measures with spot prices and option prices. Following their work, we summarize the joint dynamics of the daily spot and the option prices upon discretization as follows:
     \begin{equation}\label{models}
    \begin{split}
        &C_{t+1}-F_{t+1} = \rho_c (C_t - F_t) + \sigma_c \epsilon_t^c,\\
        &Y_{t+1} = Y_t + \left(r_t - \frac{1}{2}V_t + \psi_J^{\mathbb{Q}}(-i) + \eta_s V_t \right) \Delta + \sqrt{V_{t} \Delta} \epsilon_{t+1}^Y + J_{t+1}^Y,\\
        &V_{t+1} = V_t + \kappa (\theta - V_t)\Delta + \sigma_V \sqrt{V_t}\epsilon_{t+1}^V +J_{t+1}^V,
    \end{split}
    \end{equation}
    where $\Delta$ is the one day time interval here; $\epsilon_t^c$, $\epsilon_{t+1}^Y$ and $\epsilon_{t+1}^V$ follow the standard normal distribution (SND hereafter), $\epsilon_{t+1}^Y$ and $\epsilon_{t+1}^V$ are correlated with correlation $\rho$ and are independent from $\epsilon_t^c$; $C_t$ represents the market option price at time $t$; and $F_t$ denotes the model option price at time $t$ given by (\ref{pricing}). Following \cite{eraker2004stock}, we assume that $C_{t+1}-F_{t+1} \sim \mathbb{N} \left(\rho_c \left(C_{t}-F_{t}\right), \sigma_c^2\right)$.

    Let $\Theta$ denote the parameter vector of the models. We split the parameters into four groups, that is $\Theta = \{(\Theta^{\mathbb{P}}), (\Theta^{\mathbb{Q}}), (\Theta^{risk\,\, premia}), (\Theta^{pricing \,\,errors})\}$; the first group contains parameters under $\mathbb{P}$; the second one includes parameters that are unique under $\mathbb{Q}$; the risk premia of return and variance are in the third group while the fourth part contains parameters used to describe the option pricing errors of models. The SV model has no parameters that are unique under the risk-neutral measure, it has $\Theta = \{(\kappa, \theta, \sigma_v, \rho), (), (\eta_s, \eta_v), (\rho_c, \sigma_c)\}$; for SVJ, $\Theta = \{(\kappa, \theta, \sigma_v, \rho, \lambda, \sigma_J,\mu_J^\mathbb{P}), (\mu_J^\mathbb{Q}), (\eta_s, \eta_v), (\rho_c, \sigma_c)\}$; for SVCJ, $\Theta = \{(\kappa, \theta, \sigma_v, \rho, \lambda, \sigma_J, \rho_J, \\ \mu_J^\mathbb{P}, \mu_V), (\mu_J^\mathbb{Q}), (\eta_s, \eta_v), (\rho_c, \sigma_c)\}$; for SVVG, $\Theta = \{(\kappa, \theta, \sigma_v, \rho, \nu, \gamma^\mathbb{P}, \sigma^\mathbb{P}), (\gamma^\mathbb{Q}, \sigma^\mathbb{Q}), (\eta_s, \eta_v), (\rho_c, \sigma_c)\}$; and SVLS has $\Theta = \{(\kappa, \theta, \sigma_v, \rho, \alpha, \sigma), (), (\eta_s, \eta_v), (\rho_c, \sigma_c)\}$.

    Given the log stock prices $Y=\{Y_t\}_{t=0}^T$, the option prices $C=\{C_t\}_{t=0}^T$, the variance variables $V=\{V_t\}_{t=0}^T$, the jumps times/sizes $J=\{J_t\}_{t=0}^T$, the posterior of parameters and latent variables can be decomposed into the product of individual conditionals:
    \begin{equation}
    p(\Theta, V, J| Y, C) \propto p(Y, C, V, J, \Theta)= p(C|Y, V, \Theta)p(Y, V|J,\Theta)p(J|\Theta)p(\Theta),
    \end{equation}
    where
    \begin{equation}\label{pC}
        p(C|Y, V, \Theta) = \prod_{t=0}^{T-1}\frac{1}{\sqrt{2 \pi}\sigma_c} \exp\left\{-\frac{[(C_{t+1}-F_{t+1}) - \rho_c (C_t - F_t)]^2}{2\sigma_c^2}\right\}.
    \end{equation}
    For the SVVG model, a time-changing variable $G=\{G_t\}_{t=0}^T$, where\footnote{$\mathbb{G}$ denotes the Gamma distribution.} $G_{t+1} \sim \mathbb{G} (\frac{\Delta}{\nu},\nu)$, is introduced as a conditional latent variable on the jump process:
    \begin{equation}
    p(\Theta, V, J, G| Y, C) \propto p(Y, C, V, J, G, \Theta)= p(C|Y, V, \Theta)p(Y, V|J,\Theta)p(J|G, \Theta)p(G|\Theta)p(\Theta).
    \end{equation}
    For the SVLS model, an auxiliary variable series $U=\{U_t\}_{t=0}^T$ is added:
    \begin{equation}
    p(\Theta, V, J, U| Y, C) \propto p(Y, C, V, J, U, \Theta)= p(C|Y, V, \Theta)p(Y, V|J,\Theta)p(J,U|\Theta)p(\Theta).
    \end{equation}

    It is difficult to simulate random draws directly from the joint posterior densities of the models shown above; instead, we estimate parameters and latent variables by simulating from complete conditional distributions of each parameter and latent variable with the MCMC method. The complete conditional distributions of AJD models and L\'{e}vy processes under $\mathbb{P}$ can be found in several earlier studies, but few investigate the estimation of the parameters by a Bayesian approach under $\mathbb{Q}$. \cite{broadie2007model} simulate posterior distributions of $\Theta^{\mathbb{P}}$ with derived complete conditional distributions and then calibrate models with the estimated parameters under $\mathbb{P}$ (represented by $\widehat{\Theta^{\mathbb{P}}}$) to obtain the values of $\Theta^{\mathbb{Q}}$ based on the following objective function:
    \begin{equation}
        \Theta ^{\mathbb{Q}} = {\rm arg\, min}\sum_{t=1}^{T} \sum_{n=1}^{O_t}[IV_t(K_n, \tau_n, S_t, r_t, V_t) - IV_t(\Theta^{\mathbb{Q}} | \widehat{\Theta^{\mathbb{P}}}, K_n, \tau_n, S_t, r_t, V_t)]^2,
    \end{equation}
    here  $O_t$ is the number of options at time $t$; $S_t$, $V_t$ and $r_t$ denote the spot price, instant variance and risk-free rate at time $t$, respectively; $K_n$ and $\tau_n$ represent the strike price and expiration of the $n$-th option; and $IV$ is the implied volatility. This is a two-step estimation method. A disadvantage of this method is that only the mean values of the posterior of the parameters under $\mathbb{P}$ are considered when calibrating the models, overlooking other possible values in the posterior distribution of parameters under $\mathbb{P}$. This two-step estimation method produces an interval estimation (under $\mathbb{P}$) plus a point estimation (under $\mathbb{Q}$), which is not ideal for taking parameter estimation risk into consideration. A different approach is presented by \cite{yu2011mcmc} who derive the complete conditional distribution of each individual parameter and latent variable under both measures, enabling the simulation of  posterior samples of parameters and latent variables with the MCMC method. For $\Theta^{\mathbb{P}}$, they apply almost the same method with complete conditional distributions as \cite{broadie2007model}, after which the random draws are accepted/rejected with the Damlen, Wakefild and Walker method (\citeauthor{damlen1999gibbs}, \citeyear{damlen1999gibbs}) based on the likelihood value calculated with (\ref{pC}) whilst the Metropolis-Hasting algorithm is used to estimate $\Theta^{\mathbb{Q}}$. In the two-step method of \cite{broadie2007model}, $\Theta^{\mathbb{P}}$ are estimated from spot prices and $\Theta^{\mathbb{Q}}$ are calibrated with $\widehat{\Theta^{\mathbb{P}}}$ and option data whereas \cite{yu2011mcmc} do these in one step, and parameters are estimated with both spot price and option data.\footnote{Candidate points of $\Theta^{\mathbb{P}}$ are generated based on the posterior distribution with spot prices and only the points that also fit option prices are accepted.}

    Combining the methods of \cite{yu2011mcmc} and \cite{broadie2007model}, we estimate $\Theta^{\mathbb{P}}$ with only spot prices based on the MCMC methods introduced in \cite{li2006bayesian} while the other parameters are estimated with the method of \cite{yu2011mcmc}. Our method has less computational burden than \cite{yu2011mcmc}'s approach and, compared with \cite{broadie2007model}, our  method estimates risk-neutral parameters and real-world parameters jointly, so that the PER can be captured fully. More detailed discussions of our MCMC methods are documented in Appendix \ref{appB}.

  \section{Empirical Analysis}\label{EA}

    \subsection{Data}

    The empirical analysis is based on the S\&P 500 index spot price and the corresponding S\&P 500 index call option prices from January 3, 1996 to December 29, 2017. We follow \cite{yu2011mcmc} who choose one short-term at-the-money (ATM) call option each day. The option is required to have a time-to-expiration between 20 and 50 days and its strike to spot price ratio is closest to 1. We directly use the ATM-forward call options with 30 days to expiry as our options data. The ATM-forward option dataset is downloaded in the Std\_Option\_Price file from Option Metrics. This file contains information on ATM-forward options with expiration ranges from 30 to 730 calendar days, and the 30-days-to-expiration call options fully match \cite{yu2011mcmc}'s requirements. The forward price of underlying on the expiration date of the option is calculated with the zero-coupon yield curve and projected dividends; the strike price of the option equals the forward price; the implied volatility and premium on these standardized options are calculated daily using linear interpolation from the volatility surface, which is computed with a kernel smoothing technique. Compared with picking options in the real market, the standardized options have constant duration and they are exactly ATM, reducing measurement error arising from options that vary in maturity and moneyness. Moreover, the daily Treasury yield curve rates from the U.S. Department of the Treasury are used as the risk-free rate. The daily Treasury yield curve rates are also referred to as constant duration Treasury rates, which provide the yield curve at fixed maturities. One-month Treasury yield curve rates are not available prior to July 31, 2001, and the three-month yield curve is used before this date. More details are provided in the Supplementary Appendix.

  \subsection{Comparison of Model Parameters Inference}

In addition to the stochastic volatility models, we also use the constant volatility model with Merton jumps proposed by \cite{merton1976option} (MJD) for comparison. The SVCJ model nests MJD by setting $V_{t+1} = V_{t} = (\sigma_{MJD}^{\mathbb{Q}})^2$, where $\sigma_{MJD}^{\mathbb{Q}}$ is the constant volatility of MJD under the risk-neutral measure. We employ MCMC to estimate $\Theta = \{(\Theta^{\mathbb{P}}), (\Theta^{\mathbb{Q}}), (\Theta^{risk\,\, premia}), \\(\Theta^{pricing \,\,errors})\}$. Table \ref{esti} presents the annualized parameter estimates, specifically the estimated posterior mean of the parameters, and their corresponding standard deviation (in parenthesis).

\begin{table}[!ht]
    \begin{center}
    \scriptsize
    \setlength{\abovecaptionskip}{0pt}%
    \setlength{\belowcaptionskip}{3pt}%
    \caption{Parameter Estimates of Various Models}\label{esti}
    \setlength{\tabcolsep}{6mm}{
    \begin{tabular}{@{}lcccccc@{}}
\toprule
              & MJD      & SV       & SVJ      & SVCJ     & SVVG     & SVLS     \\ \midrule
$\sigma_{MJD}^{\mathbb{Q}}$ & 0.1149   &          &          &          &          &          \\
               & (0.0032) &          &          &          &          &          \\
$\kappa$       &          & 4.5557   & 4.2287   & 6.6665   & 3.9114   & 4.6230    \\
               &          & (0.4909) & (0.5413) & (0.7563) & (0.4472) & (0.5371) \\
$\theta$       &          & 0.0347   & 0.0331   & 0.0214   & 0.0332   & 0.0314   \\
               &          & (0.0035) & (0.0035) & (0.0024) & (0.0038) & (0.0034) \\
$\sigma_V$     &          & 0.4667   & 0.4359   & 0.4173   & 0.4595   & 0.4773   \\
               &          & (0.0155) & (0.0202) & (0.0169) & (0.0184) & (0.0137) \\
$\rho$         &          & -0.8173  & -0.7750   & -0.7642  & -0.8701  & -0.8440   \\
               &          & (0.0513) & (0.0386) & (0.0578) & (0.0344) & (0.0342) \\
$\eta_s$       & 0.0001   & 0.4667   & 0.5880    & 0.5930    & 1.0119   & 0.5591   \\
               & (0.0467) & (0.0155) & (2.8090)  & (2.0234) & (2.9937) & (1.0613) \\
$\eta_v$       &          & -19.8169 & -16.4552 & -17.7829 & -16.3290  & -17.8977 \\
               &          & (0.9664) & (1.2666) & (1.2619) & (1.0617) & (1.1024) \\
$\rho_c$       & 0.9757   & 0.9600     & 0.9163   & 0.9198   & 0.9651   & 0.9339   \\
               & (0.0031) & (0.0045) & (0.0155) & (0.0133) & (0.0115) & (0.0077) \\
$\sigma_c$     & 2.3870    & 2.8215   & 2.6652   & 2.7076   & 2.5980    & 2.6708   \\
               & (0.0233) & (0.0381) & (0.0479) & (0.0424) & (0.0293) & (0.0366) \\
$\lambda$      & 54.1371  &          & 2.1108   & 2.1754   &          &          \\
               & (4.9170)  &          & (1.1312) & (0.7299) &          &          \\
$\mu_J^\mathbb{P}$      & -0.0024  &          & -0.0120   & -0.0138  &          &          \\
               & (0.0008) &          & (0.0040)  & (0.0058) &          &          \\
$\mu_J^\mathbb{Q}$      & -0.0003  &          & -0.0872  & -0.1150   &          &          \\
               & (0.0009) &          & (0.0181) & (0.0154) &          &          \\
$\sigma_J$     & 0.0204   &          & 0.0184   & 0.0173   &          &          \\
               & (0.0008) &          & (0.0038) & (0.0029) &          &          \\
$\mu_V$        &          &          &          & 0.0401   &          &          \\
               &          &          &          & (0.0001) &          &          \\
$\rho_J$       &          &          &          & -0.0023  &          &          \\
               &          &          &          & (0.0011) &          &          \\
$\nu$          &          &          &          &          & 0.0264   &          \\
               &          &          &          &          & (0.0149) &          \\
$\gamma^\mathbb{P}$     &          &          &          &          & -0.1559  &          \\
               &          &          &          &          & (0.0573) &          \\
$\sigma^\mathbb{P}$     &          &          &          &          & 0.2152   &          \\
               &          &          &          &          & (0.0319) &          \\
$\gamma^\mathbb{Q}$     &          &          &          &          & -0.3185  &          \\
               &          &          &          &          & (0.0863) &          \\
$\sigma^\mathbb{Q}$     &          &          &          &          & 1.2652   &          \\
               &          &          &          &          & (0.0702) &          \\
$\alpha$       &          &          &          &          &          & 1.9268   \\
               &          &          &          &          &          & (0.0050)  \\
$\sigma$       &          &          &          &          &          & 0.4800     \\
               &          &          &          &          &          & (0.0261) \\  \midrule
 DIC          & 12664.68 & 14573.85 & 14405.48 & 14367.40 & 14153.17 & 14167.09 \\\bottomrule
\end{tabular}}
    \end{center}
 \vspace{-0.45cm}
    \begin{tablenotes}
      \scriptsize
      \item NOTE: Estimates of parameters (with standard errors in parenthesis) using daily spot price on the S\&P 500 and daily price of standardized at-the-money-forward call options with 30 days to expiration between January 3, 1996 and December 29, 2017. The first 10000 runs are discarded as ``burn-in" period and the last 20000 iterations in the MCMC simulations are used to estimate the parameters.
    \end{tablenotes}
    \end{table}
The estimated volatility of MJD ($\sigma_{MJD}^{\mathbb{Q}}$) is 0.1149. The estimates of parameters related to the variance process of stochastic volatility models are relatively similar. The reversion speed, $\kappa$, ranges from 3.9114 for SVVG to 6.6665 for SVCJ, the faster reversion speed of SVCJ can be explained by the jumps in the variance process, which require a higher reversion speed to drive the variance to the long-run mean level. The estimates of the average variance for SV, SVJ, SVVG and SVLS are $\theta$, which are approximately 0.033; the long-run mean-variance level of SVCJ is $\theta + \mu_V \lambda / \kappa = 0.0345$. The estimates for $\sigma_V$ and $\rho$ are also quite close for all models and consistent with existing studies. The estimated jump intensity, $\lambda$, of MJD is 54.1371, indicating about 54 price jumps per annum with very small mean jump sizes (-0.0024 for $\mu_J^{\mathbb{P}}$ and almost 0 for $\mu_J^{\mathbb{Q}}$). By contrast, jumps in price are infrequent events in SVJ and SVCJ (both reveal about two jumps per year). The mean jump sizes for both SVJ and SVCJ under the real-world measure ($\mu_J^{\mathbb{P}}$) are negative and the mean jump sizes under the risk-neutral measure ($\mu_J^{\mathbb{Q}}$) have larger absolute values. SVJ and SVCJ capture the price movements with infrequent large jumps while L\'{e}vy jump models are able to capture both infrequent large jumps and frequent small jumps. According to our results, the jump distribution under the physical measure for SVVG is negatively skewed, $\gamma^{\mathbb{P}}$ is -0.1559, while the risk-neutral jump distribution is also negatively skewed with even larger jump sizes.

    \begin{table}[htbp]
    \begin{center}
    \scriptsize
    \setlength{\abovecaptionskip}{0pt}%
    \setlength{\belowcaptionskip}{-5pt}%
    \caption{KS and Abadie's Test Results}\label{ks_abadie}
    \setlength{\tabcolsep}{5.5mm}{
   \begin{tabular}{@{}lllllll@{}}
\toprule
                          & MJD    & SV     & SVJ    & SVCJ   & SVVG   & SVLS   \\ \midrule
Panel A. Return Residuals &        &        &        &        &        &        \\ \midrule
KS statistics             & 0.0419 & 0.0362 & 0.0321 & 0.0291 & 0.0453 & 0.0493 \\
KS p-values               & 0.0000 & 0.0000 & 0.0000 & 0.0002 & 0.0000 & 0.0000 \\
Abadie p-values           & 0.0002 & 0.0062 & 0.0193 & 0.0384 & 0.0003 & 0.0001 \\ \midrule
Panel B. Volatility Residuals                                                  \\ \midrule
KS statistics             &        & 0.0319 & 0.0378 & 0.0289 & 0.0294 & 0.0226 \\
KS p-values               &        & 0.0000 & 0.0000 & 0.0002 & 0.0001 & 0.0068 \\
Abadie p-values           &        & 0.0183 & 0.0035 & 0.0213 & 0.0299 & 0.0990 \\ \bottomrule
\end{tabular}}
    \end{center}
 \vspace{-0.45cm}
    \begin{tablenotes}
      \scriptsize
      \item NOTE: This table provides the KS statistics and corresponding p-values for return residuals (for all models) and for volatility residuals (of stochastic volatility models). The p-values calculated with Abadie's bootstrap method are also reported.
    \end{tablenotes}
    \end{table}
For a goodness-of-fit comparative analysis, we apply the Kolmogorov-Smirnov (KS) test with the improved adjustment of \cite{abadie2002bootstrap} to the return residuals (for all models) and the volatility residuals (for stochastic volatility models). The results are reported in Table \ref{ks_abadie}. According to these, the KS tests reject the null hypothesis that the residuals of all models follow the SND; by contrast, Abadie's test fails to reject the null hypothesis for the volatility residuals of SVLS. Abadie's p-value of the SVCJ return residuals is relatively large at 0.0384, indicating that the specification of jumps in volatility improves modeling performance. 
   
Moreover, we apply the deviance information criterion (DIC) (\citeauthor{spiegelhalter2002bayesian}, \citeyear{spiegelhalter2002bayesian} and \citeauthor{berg2004deviance}, \citeyear{berg2004deviance}) and Bayes factors (\citeauthor{kass1995bayes}, \citeyear{kass1995bayes} and \citeauthor{chib2002markov}, \citeyear{chib2002markov}) to compare models from a Bayesian inference perspective.\footnote{For the definition of the DIC and Bayes factors see the Supplementary Appendix.} The DIC values are shown in the last row of Table \ref{esti}; a smaller DIC value indicates a better fit of the model to the index returns. MJD has the lowest DIC value because it generates many jumps (over 54 jumps per year) to capture the abnormal movements of market prices. For stochastic volatility models, SV is outperformed by all other stochastic volatility models with jumps. This is consistent with \cite{chernov2003alternative}; the authors state the tradeoffs among various model specifications and find that stochastic volatility and jumps in returns are crucial specifications for affine models. Except for MJD and SV, the L\'{e}vy jump models (SVVG and SVLS) have the smallest DIC values, followed by SVCJ and SVJ. Table \ref{bayes_factor} reports the log values of the Bayes factors (log-BF) of the SVLS, SVVG, SVCJ, SVJ and SV models over the MJD, SV, SVJ, SVCJ and SVVG models, respectively. A negative log-BF indicates the underperformance of the column model. Based on the log-BF values, MJD outperforms all stochastic volatility models; the SV has the worst performance while the SVVG performs best in the stochastic volatility model class. The DIC value and Bayes factors tend to prefer models with more jumps which explain abnormal market movements.
\begin{table}[!ht]
    \begin{center}
    \scriptsize
    \setlength{\abovecaptionskip}{0pt}%
    \setlength{\belowcaptionskip}{3pt}%
    \caption{Log Values of Bayes Factors of Models}\label{bayes_factor}
    \setlength{\tabcolsep}{7.5mm}{
   \begin{tabular}{@{}lrrrrr@{}}
\toprule
     & SVLS     & SVVG     & SVCJ     & SVJ      & SV       \\ \midrule
MJD  & -791.75 & -722.04 & -1008.28 & -1062.31 & -1152.43 \\
SV   & 360.69  & 430.39 & 144.14 & 90.12  &          \\
SVJ  & 270.57 & 340.27 & 54.03    &          &          \\
SVCJ & 216.54 & 286.24 &          &          &          \\
SVVG & -69.71 &          &          &          &          \\ \bottomrule
\end{tabular}}
\end{center}
\vspace{-0.45cm}
    \begin{tablenotes}
      \scriptsize
      \item NOTE: Log values of the Bayes factors of the SVLS, SVVG, SVCJ, SVJ and SV models over the MJD, SV, SVJ, SVCJ and SVVG models, respectively.
    \end{tablenotes}
    \end{table}
\subsection{Model Risk} \label{model-risk}
    
\begin{figure}[ht]
    \begin{center}
    \includegraphics [width=\linewidth, height = \textheight, keepaspectratio]{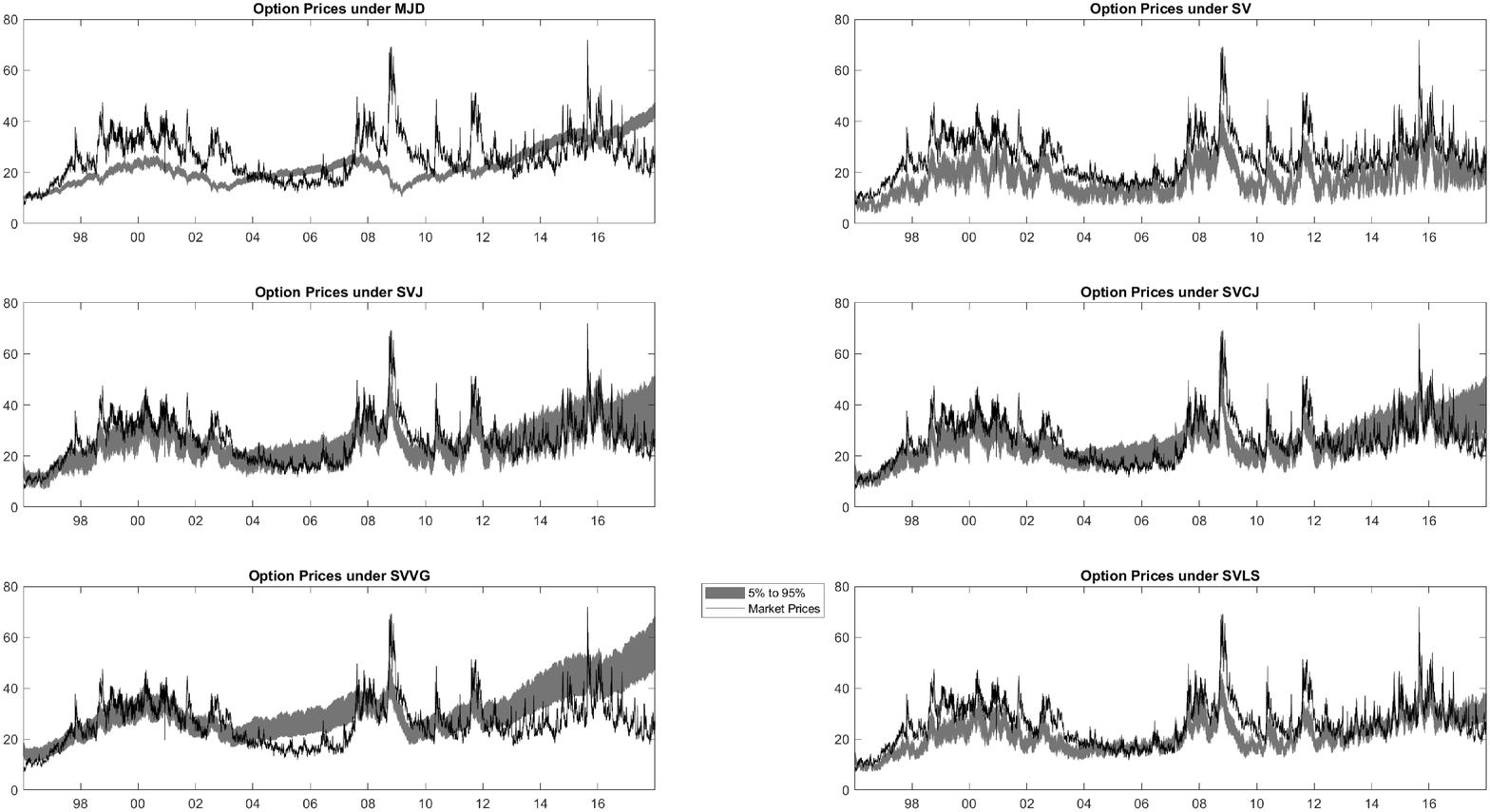}
    \caption{\small{Call Option Prices under Various Models.\newline This figure presents the 5\% to 95\% quantiles of estimated prices distribution; the black line is the market price of options as provided by Option Metrics. The results are based on daily spot prices on the S\&P 500 and daily prices of standardized at-the-money-forward call options with 30 days to expiration between January 3, 1996 and December 29, 2017. Estimated prices are updated within the estimation process at the end of each iteration after burn-in.}}\label{epp}
    \end{center}
    \end{figure}

Figure \ref{epp} shows the 5\% to 95\% quantiles of estimated prices, after burn-in during the estimation process for all models. The MJD model cannot accurately capture the dynamics of option prices due to the inherent inability of MJD in modeling time-varying variance processes. By contrast, the estimated prices of all stochastic models have a similar trend to the market prices. For stochastic volatility, models with jumps can capture the volatile option prices well, as the peaks of market prices are almost all in the 5\% to 95\% quantile of estimated prices of these models. Most of market prices lie in the estimated prices range of SVJ due to wide credibility intervals. Although with a narrower range, the estimated prices of SVCJ also cover the market prices well, except for the apparent mispricing before 2010. SVVG has excellent performance during market turmoil; however, it fails to capture the market dynamics during tranquil periods (the period after 2004 until the financial crisis and the period after 2016); by contrast, SVLS performs well during these tranquil periods. Besides, SV reveals a comparatively better pricing ability in 2017.

  The model risk and pricing performance of models are reported in Table \ref{result}. TMR, PER and MSR are defined in (\ref{tomr3}), (\ref{perre}) and (\ref{msrdcrother}), respectively. According to Panel A, SV has the largest TMR, followed by SVVG and MJD. However, the dominant model risk of SVVG is PER while the dominant model risk of SV and MJD is MSR. SVLS has the lowest TMR, followed by SVCJ and SVJ. The TMR of these two AJD models mostly consists of PER; moreover, SVCJ has lower PER and greater MSR than SVJ. 
  Panel B reports the mean values of the percentage model risk, which are calculated as the values of model risk divided by option prices. Panel C reports the PER of long and short positions; this shows that the mean values of PER of long positions are lower than those of short positions for L\'{e}vy jump models and SV, while the PER of a long position for AJD models and MJD is slightly higher than that of a short position. As reported in Panel D, AJD models have the lowest pricing errors (PE, APE and APE(\%)); followed by L\'{e}vy jump models and MJD. SV reveals the worst pricing performance.
    
\begin{table}[!ht]
    \begin{center}
    \scriptsize
    \caption{Model Risk and Pricing Performance of Models}\label{result}
    \setlength{\tabcolsep}{7mm}{
   \begin{tabular}{@{}lllllll@{}}
\toprule
       & SVLS      & SVVG      & SVCJ      & SVJ       & SV        & MJD       \\ \midrule
\multicolumn{7}{l}{Panel A. Model Risk of Models}                              \\ \midrule
TMR     & 6.0562  & 8.3272   & 6.2299 & 6.3251 & 8.7559 & 8.2853 \\
PER     & 2.7469  & 4.5345   & 4.6207 & 4.9616 & 3.2687 & 1.4138 \\
MSR     & 3.3092  & 3.7927   & 1.6092 & 1.3635 & 5.4872 & 6.8715 \\ \midrule
\multicolumn{7}{l}{Panel B. Model Risk of Models (\%)}                \\ \midrule
TMR(\%)    & 21.81\% & 35.15\% & 24.15\% & 24.71\% & 32.57\% & 29.18\% \\
PER(\%)    & 10.73\% & 18.74\% & 18.80\% & 20.39\% & 12.91\% & 5.77\% \\
MSR(\%)    &11.07\% & 16.41\% & 5.34\%  & 4.32\%  & 19.66\% & 23.41\% \\ \midrule
\multicolumn{7}{l}{Panel C. PER of Long and Short Positions}               \\ \midrule
Long    & 2.5957  & 4.1590    & 4.6669 & 4.9995 & 3.0368 & 1.4255 \\
Short   & 2.8982  & 4.9100     & 4.5746 & 4.9237 & 3.5006 & 1.4021 \\ \midrule
\multicolumn{7}{l}{Panel D. Pricing Performance}                               \\ \midrule
PE       & 4.6671  & -3.8331  & 0.9257 & 1.0386 & 8.8167 & 4.0031 \\
APE      & 5.7338  & 7.4055   & 5.1075 & 4.8393 & 8.8407 & 8.2018 \\
APE(\%)     & 20.22\%  & 31.47\%   & 19.50\%  & 18.43\% & 32.79\% & 28.81\% \\ \bottomrule
\end{tabular}}
    \end{center}
 \vspace{-0.45cm}
    \begin{tablenotes}
      \scriptsize
      \item NOTE: Results are based on the daily spot price on the S\&P 500 and daily price of standardized at-the-money-forward call options with 30 days to expiry between January 3, 1996 and December 29, 2017. Panel A presents mean values of model risk; Panel B shows the mean values of the percentage model risk, calculated as model risk over option price; Panel C displays PER of long and short positions; Panel D presents pricing performance, where PE denotes pricing error, APE represents absolute pricing error, and APE(\%) is the percentage absolute error.
    \end{tablenotes}
    \end{table}

    In order to analyze the model risk of models under different market periods, we split the sample period into seven time windows by detecting  abrupt changes in the implied volatility of the standardized ATM options based on the method proposed by \cite{killick2012optimal}. Using  average values of implied volatility, we identify the following time windows:
     period \raisebox{.5pt}{\textcircled{\raisebox{-.9pt} {1}}}: January 3, 1996 to June 4, 1997;
       period \raisebox{.5pt}{\textcircled{\raisebox{-.9pt} {2}}}: June 5, 1997 to October 13, 2003;
     period \raisebox{.5pt}{\textcircled{\raisebox{-.9pt} {3}}}: October 14, 2003 to July 24, 2007;
     period \raisebox{.5pt}{\textcircled{\raisebox{-.9pt} {4}}}: July 25, 2007 to September 23, 2008;
      period \raisebox{.5pt}{\textcircled{\raisebox{-.9pt} {5}}}: September 24, 2008 to May 1, 2009;
       period \raisebox{.5pt}{\textcircled{\raisebox{-.9pt} {6}}}: May 4, 2009 to June 29, 2012;
     and period \raisebox{.5pt}{\textcircled{\raisebox{-.9pt} {7}}}: July 2, 2012 to December 29, 2017.

    The values of the mean and standard deviation of the implied volatility for the standardized ATM options during these seven time windows are reported in Panel A of Table \ref{ivsta}. Period \raisebox{.5pt}{\textcircled{\raisebox{-.9pt} {3}}} is the most tranquil period, followed by periods \raisebox{.5pt}{\textcircled{\raisebox{-.9pt} {7}}} and \raisebox{.5pt}{\textcircled{\raisebox{-.9pt} {1}}}, whose mean implied volatility values are all below 0.2. During these three periods the market is stable. In contrast, the market is turbulent  during periods \raisebox{.5pt}{\textcircled{\raisebox{-.9pt} {6}}}, \raisebox{.5pt}{\textcircled{\raisebox{-.9pt} {4}}} and \raisebox{.5pt}{\textcircled{\raisebox{-.9pt} {2}}}. Period  \raisebox{.5pt}{\textcircled{\raisebox{-.9pt} {2}}} is very long and covers the 1997 Asian financial crisis, the Russian financial crisis that hit on August 17, 1998, the Brazilian currency crisis from 1998 to 1999, the  dot-com crash from March 11, 2000, to October 9, 2002, the 1998-to-2002 Argentine great depression, the 911 (September 11, 2001) and the WorldCom accounting scandal in 2002; period \raisebox{.5pt}{\textcircled{\raisebox{-.9pt} {4}}} involves the subprime crisis; and period \raisebox{.5pt}{\textcircled{\raisebox{-.9pt} {6}}} captures the European credit crisis since the end of 2009. Period \raisebox{.5pt}{\textcircled{\raisebox{-.9pt} {5}}} marks the well-know Financial Crisis of 2008, when the mean implied volatility reaches 0.4486 with a standard deviation of 0.0966.

   \begin{table}[htbp]
    \centering
    \scriptsize
    \setlength{\abovecaptionskip}{0pt}%
    \setlength{\belowcaptionskip}{3pt}%
    \caption{Statistics on Implied Volatility and the Differences between the PER of Long and Short Positions}\label{ivsta}
    \setlength{\tabcolsep}{4.1mm}{
   \begin{tabular}{@{}llllllll@{}}
\toprule
Period & \raisebox{.5pt}{\textcircled{\raisebox{-.9pt} {1}}} & \raisebox{.5pt}{\textcircled{\raisebox{-.9pt} {2}}} & \raisebox{.5pt}{\textcircled{\raisebox{-.9pt} {3}}} & \raisebox{.5pt}{\textcircled{\raisebox{-.9pt} {4}}} & \raisebox{.5pt}{\textcircled{\raisebox{-.9pt} {5}}} & \raisebox{.5pt}{\textcircled{\raisebox{-.9pt} {6}}} & \raisebox{.5pt}{\textcircled{\raisebox{-.9pt} {7}}} \\ \midrule
\multicolumn{8}{l}{Panel A. The Mean and Standard Deviation of the Implied Volatility of the Standardized ATM Options}                              \\ \midrule
Mean    & 0.1568                           & 0.2265                           & 0.1244                           & 0.2182                           & 0.4486                           & 0.2080                           & 0.1272                           \\
Std     & 0.0236                           & 0.0484                           & 0.0207                           & 0.0351                           & 0.0966                           & 0.0557                           & 0.0336                           \\ \midrule
\multicolumn{8}{l}{Panel B. Mean Differences between the PER of Long and Short Positions}                              \\ \midrule
MJD    &  {0.0122**}  &  {0.0199**}  &  {0.0215**}  &  {0.0240**}  &  {0.0151**}  &  {0.0205**}  &  {0.0342**}  \\
SV     &  {-0.2101**} &  {-0.3627**} &  {-0.4606**} &  {-0.3355**} &  {-0.3412**} &  {-0.4233**} &  {-0.7126**} \\
SVJ    &  {0.0568**}  &  {-0.0224*} &  {0.1729**}  &  {-0.0715**} &  {0.1547**}  & -0.0126          &  {0.2006**}  \\
SVCJ   &  {0.1331**}  &  {-0.1416**} &  {0.2406**}  &  {-0.4108**} & 0.0108           & -0.0074          &  {0.4233**}  \\
SVVG   &  {-0.4528**} &  {-0.5240**} &  {-0.8259**} &  {-0.3697**} &  {-0.3366**} &  {-0.7433**} &  {-1.1700**} \\
SVLS   &  {-0.1368**} &  {-0.3580**} &  {-0.1373**} &  {-0.7723**} &  {-0.3568**} &  {-0.3039**} &  {-0.2879**}  \\ \bottomrule
\end{tabular}}
\begin{tablenotes}
      \scriptsize
      \item NOTE: Panel A reports the mean and standard deviation of the implied volatility of the Standardized ATM options in different periods; Panel B reports the mean values of the differences between the PER of long and short positions for all models under different time windows.
      * and ** indicate values significant at 5\% and 1\% significance levels, respectively.
    \end{tablenotes}
    \end{table}

\begin{figure}[!ht]
    \begin{center}
    \includegraphics [width=\linewidth, height = \textheight, keepaspectratio]{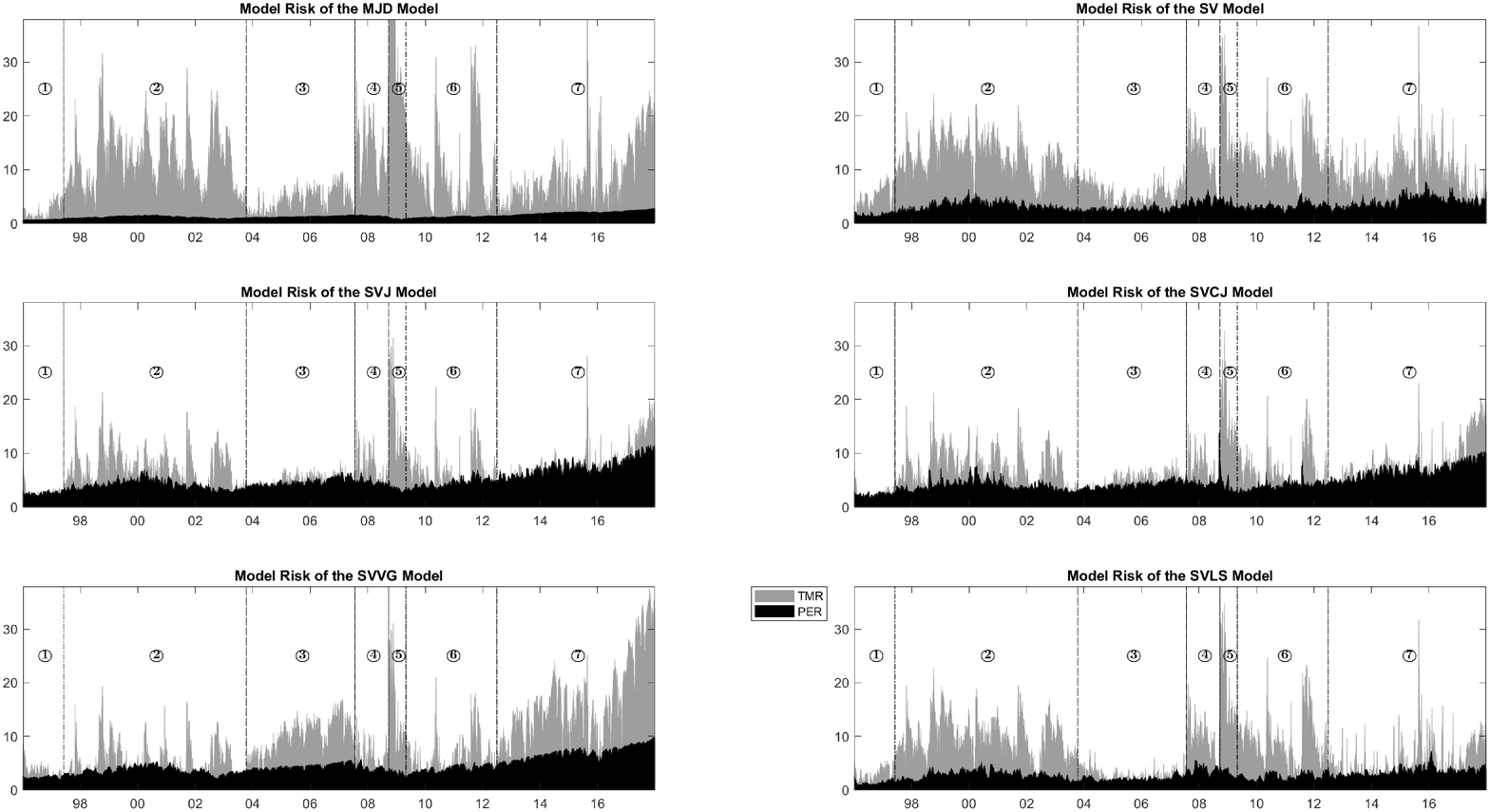}
    \caption{\small{Model Risk of Models.\newline This figure presents the TMR (grey columns) and PER (black columns) of models. The results are based on the daily spot price on the S\&P 500 and daily price of standardized at-the-money-forward call options with 30 days to expiration between January 3, 1996 and December 29, 2017.}}\label{mrrr}
    \end{center}
    \end{figure}

Figure \ref{mrrr} illustrates the model risk estimates of our models. The black columns mark the PER of models while the grey columns are the values of the TMR of models; the differences between grey and black columns being the MSR. The MJD has the smallest PER and remarkably high MSR. SV reveals very large MSR during volatile periods \raisebox{.5pt}{\textcircled{\raisebox{-.9pt} {2}}}, \raisebox{.5pt}{\textcircled{\raisebox{-.9pt} {4}}} and \raisebox{.5pt}{\textcircled{\raisebox{-.9pt} {6}}}, and relatively small MSR during tranquil periods \raisebox{.5pt}{\textcircled{\raisebox{-.9pt} {1}}}, \raisebox{.5pt}{\textcircled{\raisebox{-.9pt} {3}}} and \raisebox{.5pt}{\textcircled{\raisebox{-.9pt} {7}}}. Stochastic volatility models with jumps have a considerably smaller MSR risk during market turmoil (periods \raisebox{.5pt}{\textcircled{\raisebox{-.9pt} {2}}}, \raisebox{.5pt}{\textcircled{\raisebox{-.9pt} {4}}} and \raisebox{.5pt}{\textcircled{\raisebox{-.9pt} {6}}}) compared with SV. Besides, AJD models and SVLS also have minor MSR in stable periods. By contrast, SVVG reveals large MSR during the tranquil periods \raisebox{.5pt}{\textcircled{\raisebox{-.9pt} {3}}} and \raisebox{.5pt}{\textcircled{\raisebox{-.9pt} {7}}}. However, during  periods \raisebox{.5pt}{\textcircled{\raisebox{-.9pt} {2}}}, \raisebox{.5pt}{\textcircled{\raisebox{-.9pt} {4}}} and \raisebox{.5pt}{\textcircled{\raisebox{-.9pt} {6}}}, characterized by a turbulent market, SVVG has less MSR compared with other stochastic volatility models. These findings support adding jumps to stochastic volatility models, especially when the market is volatile.
All models reveal substantial model risk during period \raisebox{.5pt}{\textcircled{\raisebox{-.9pt} {5}}}, the Financial Crisis. The SV model, with no jumps, has the worst performance compared with other stochastic volatility models with jumps. It is worth noting that the sizes of the MSR of all models with jumps reveal a distinct spike at the end of the research period, while the SV performs very well during this period when the S\&P 500 index rises considerably in 2017 with few big swings.\footnote{More results are available in the Supplementary Appendix.}

  Panel B of Table \ref{ivsta} reports the means of the differences between the PER of long and short positions in different periods.\footnote{The figure of the PER of long and short positions of models can be found in the Supplementary Appendix.} For the MJD model, the long position always exhibits a significantly higher PER than the short position; by contrast, for stochastic volatility models, a short position tends to have a higher PER during turbulent periods \raisebox{.5pt}{\textcircled{\raisebox{-.9pt} {2}}}, \raisebox{.5pt}{\textcircled{\raisebox{-.9pt} {4}}} and \raisebox{.5pt}{\textcircled{\raisebox{-.9pt} {6}}}, although the difference is not significant for AJD models in period \raisebox{.5pt}{\textcircled{\raisebox{-.9pt} {6}}}. A short position always reveals a higher PER for L\'{e}vy jump models. By contrast, a long position is more likely to bear higher model risk for AJD models.
   L\'{e}vy jump models and SV reveal larger differences between long and short positions compared with AJD models and MJD.

    \subsection{Model Comparison in terms of Model Risk}
We examine the performance of models in capturing the risk-neutral dynamics by testing the null hypothesis that the squared pricing errors of the two models are equal, using the \cite{diebold1995paring} (DM) tests for non-nested models and the \cite{clark2007approximately} (CW) tests for nested models.\footnote{See the Supplementary Appendix for a presentation of the DM and CW tests.} Table \ref{DM_tests} reports DM and CW statistics for squared option pricing errors in the Squared Pricing Error panel. Rejections of the null hypothesis are denoted using stars; for the DM tests, a positive (negative) value indicates that the column model has significantly larger (smaller) squared pricing errors than the corresponding row model.
    \begin{table}[htbp]
    \centering
    \scriptsize
    \setlength{\abovecaptionskip}{0pt}%
    \setlength{\belowcaptionskip}{3pt}%
    \caption{Model Comparison}\label{DM_tests}
    \setlength{\tabcolsep}{0.7mm}{
  \begin{tabular}{@{}lllllllllllllll@{}}
\toprule
     & \multicolumn{4}{l}{Squared Pricing Errors}                               &  & \multicolumn{4}{l}{MSR(\%)}                                        &  & \multicolumn{4}{l}{MSR}                                         \\ \midrule
\multicolumn{15}{l}{Panel A. DM statistics}                                                                                                                                                                      \\ \midrule
     & SVLS             & SVVG             & SV               &        &  & SVLS             & SVVG             & SV               &        &  & SVLS             & SVVG             & SV               &        \\ \midrule
SVVG & -0.9785          &                  &                   &                 &  & -1.6237          &                 &                   &                 &  & -0.7288          &                  &                  &        \\
SVCJ & 1.3130           & 1.5761           &                   &                 &  &  {3.5859**}  &  {2.2955*} &                   &                 &  &  {5.4824**}  &  {1.8497*}  &                  &        \\
SVJ  &  {1.8409*}  & 1.6051           &                   &                 &  &  {4.6174**}  &  {2.2979*} &                   &                 &  &  {4.6850**}  &  {1.8859*}  &                  &        \\
SV   &  {-8.2820**} & 0.0341           &                   &                 &  &  {-8.1634**} & 0.7090          &                   &                 &  &  {-8.4180**} & -0.3409          &                  &        \\
MJD  &  {-3.5977**} & -0.4073 & 
-0.9574 &                 &  &  {-4.7764**} & -0.3775         &  {-2.1789*} &                 &  &  {-3.8131**} &  {-2.0102*} &  {-2.5578**} &        \\ \midrule
\multicolumn{15}{l}{Panel B. CW statistics}                                                                                                                                                                                         \\ \midrule
     & SVCJ             & SVJ              & SVLS              & SVVG            &  & SVCJ             & SVJ             & SVLS              & SVVG            &  & SVCJ             & SVJ              & SVLS             & SVVG   \\ \midrule
SVJ  &  {3.7216**}  &                  &                   &                 &  &  {2.7498**}  &                 &                   &                 &  &  {3.3602**}  &                  &                  &        \\
SV   & 0.6632           & 0.4290           & -2.8533           &  {1.7723*} &  & -0.7047          & -2.0032         & -3.2299           &  {2.1046*} &  & -3.3965          & -3.5998          & -4.5702          & 1.5062 \\
MJD  & -4.1330          & -4.1802          &                   &                 &  & -4.4129          & -4.3658         &                   &                 &  & -3.6706          & -3.3786          &                  &        \\ \bottomrule
\end{tabular}}
    \begin{tablenotes}
      \scriptsize
      \item NOTE: Panel A reports DM statistics for squared pricing errors, MSR(\%), and MSR of SVLS, SVVG and SV to other corresponding non-nested models. * and ** indicate values significant at 10\% and 2\% significance levels for the two-sided test, respectively;  Panel B reports the CW statistics of SVCJ, SVJ, SVLS and SVVG to other corresponding nested models. * and ** indicate values significant at 5\% and 1\% significance levels for the one-sided test, respectively.
    \end{tablenotes}
    \end{table}
According to the DM tests, although the MJD model has excellent ability to capture index dynamics under the physical measure, it fails to capture the risk-neutral dynamics. The squared pricing errors of MJD are significantly greater than those of SVLS; moreover, SVLS reveals significantly greater squared pricing errors than SVJ. In addition, we test the null hypothesis that the MSR(\%) and MSR of the two models are equal. MJD has significantly higher MSR(\%) than SVLS and SV; as well as greater MSR compared with SVLS, SVVG and SV. The MSR(\%) and MSR of AJD models are significantly less than those of L\'{e}vy jump models. SV has significantly greater squared pricing errors, MSR(\%) and MSR than SVLS.
For the CW tests, based on Panel B, SVCJ is rejected compared with SVJ in all terms. This indicates that there is no evidence of jumps in the volatility process. Moreover, SVVG is also rejected compared with SV in terms of squared pricing errors and MSR(\%).

\section{Regression Analysis}\label{ra}

In this section, we highlight the necessity of measuring the PER and MSR separately in terms of explaining absolute pricing errors. We also consider forecasting model risk using market data.

Let $\epsilon_t( \mathcal{H};\mathcal{M}(\Theta), \mathcal{D}, \mathcal{K}) = |\widehat{F_t}( \mathcal{H};\mathcal{M}(\Theta), \mathcal{D}, \mathcal{K}) - C_t(\mathcal{H})|$ represent the absolute pricing error of option $\mathcal{H}$ conditional on model $\mathcal{M}$ with the parameter vector $\Theta$, observed dataset $\mathcal{D}$, and methodology $\mathcal{K}$ at time $t$. We run the following regression to explore the explanatory power of the PER and MSR in explaining absolute pricing errors.
      \begin{equation}\label{reg_mrr}
      \epsilon_t( \mathcal{H};\mathcal{M}(\Theta), \mathcal{D}, \mathcal{K}) = \beta_0 + \beta_1 \rho^{PER}_{\eta, \,t}(\mathcal{H};\mathcal{M}(\Theta), \mathcal{D}, \mathcal{K}) + \beta_2 \rho^{MSR}_{\eta, \,t}( \mathcal{H};\mathcal{M}(\Theta), \mathcal{D}, \mathcal{K}) + \varepsilon_t.
      \end{equation}
Here, we focus on testing whether $\beta_1 = \beta_2$, as this would prove that it is not necessary to separate the PER and MSR from TMR. Taking $\beta_1 - \beta_2 = \alpha$ and using (\ref{msrdcrother}), (\ref{reg_mrr}) can be rewritten as:
      \begin{equation}\label{reg_mrr1}
    \epsilon_t( \mathcal{H};\mathcal{M}(\Theta), \mathcal{D}, \mathcal{K})   = \beta_0 + \alpha \rho^{PER}_{\eta, \,t}(\mathcal{H};\mathcal{M}(\Theta), \mathcal{D}, \mathcal{K}) + \beta_2 \rho^{TMR}_{\eta, \,t}(\mathcal{H};\mathcal{M}(\Theta), \mathcal{D}, \mathcal{K}) + \varepsilon_t.
      \end{equation}
A test of $\alpha=0$ in (\ref{reg_mrr1}) is a test of $\beta_1 = \beta_2$ in (\ref{reg_mrr}). The regression results are reported in Table \ref{reg_per_tmr}. All $\alpha$'s are statistically significant, which supports the necessity of measuring PER and MSR separately.\footnote{More results are reported  in the Supplementary Appendix.}
\begin{table}[htbp]
    \begin{center}
    \scriptsize
    \setlength{\abovecaptionskip}{0pt}%
    \setlength{\belowcaptionskip}{3pt}%
    \caption{Regression Analysis with PER and TMR}\label{reg_per_tmr}
 \setlength{\tabcolsep}{7.8mm}{
    \begin{tabular}{@{}lllll@{}}
\toprule
     & C              & PER            & TMR   & Adj. $R^2$ \\ \midrule
MJD  &  {-0.11**} &  {-0.04**} &  {1.01**} & 1                         \\
SV   &  {-0.10**}  &  {-0.04**} &  {1.04**} & 0.98                      \\
SVJ  &  {-0.68**} &  {-0.48**} &  {1.25**} & 0.84                      \\
SVCJ &  {-0.66**} &  {-0.41**} &  {1.23**} & 0.87                      \\
SVVG &  {-0.10*}  &  {-0.44**} &  {1.14**} & 0.98                      \\
SVLS &  {-0.38**} &  {-0.21**} &  {1.11**} & 0.97                      \\  \bottomrule
\end{tabular}}
\end{center}
\vspace{-0.45cm}
    \begin{tablenotes}
      \scriptsize
      \item NOTE: The regression results are based on Equation (\ref{reg_mrr1}). * and ** indicate values significant at 5\% and 1\% significance levels, respectively.
    \end{tablenotes}
    \end{table}

To understand the structure of model risk, we appeal to a forecasting regression to investigate the association between model risk and several market risk factors. The regression is based on:
      \begin{equation}\label{reg_fore}
      \begin{split}
          \rho^{MR}_{\eta, \,t}( \mathcal{H};\mathcal{M}(\Theta), \mathcal{D}, \mathcal{K}) & =\beta_0 + \beta_1 IV_{t-1} + \beta_2 C\_PV_{t-1} + \beta_3 DDelta_{t-1} \\
           &+ \beta_4 Gamma_{t-1}  + \beta_5 Theta_{t-1} + \beta_6 H\_L_{t-1}\\
           & + \beta_7 DSPXV_{t-1} + \beta_8 C\_PP_{t-1} + \varepsilon_t,
      \end{split}
      \end{equation}
\noindent 
$\rho^{MR}$ can be PER, MSR and L\_S, where L\_S denotes the difference between the PER of long and short positions. $IV$ is the implied volatility of 30-days-to-expiration ATM-forward call options of the S\&P 500 Index; $C\_PV$ is the difference between daily trading volumes of call options and put options on the S\&P 500 Index over 100,000; $DDelta_t = Delta_t - Delta _{t-1}$; and $Delta$, $Gamma$ and $Theta$ are the ``greek'' sensitivities associated with option prices of 30-days-to-expiration ATM-forward call options on the S\&P 500 Index, the sizes of $Theta$ are adjusted by dividing by 100; $H\_L$ is the difference between S\&P 500 Index High price and Low price; $DSPXV_t=SPXV_t-SPXV_{t-1}$, and $SPXV$ is the trading volume of S\&P 500 Index over 100,000,000; $C\_PP$ is the difference between 30-days-to-expiration ATM-forward call option price and put option price of the S\&P 500 Index.\footnote{Based on the Dickey-Fuller test, the raw series of $Delta$ and $SPXV$ are non-stationary, so we compute the differences between consecutive observations of $Delta$ and $SPXV$; all independent variables in (\ref{reg_fore}) are stationary. We also assess the strength of collinearity among independent variables in (\ref{reg_fore}) with Belsley collinearity diagnostics, the largest condition index is 11.7834, which does not exceed the tolerance, 30. Thus, there is no evidence of multicollinearity.}
 
Regression results are presented in Table \ref{fore_mr}. The predictive power of these variables for PER is quite high for all models. The adjusted $R^2$ of the MJD PER reaches 80\%, followed by SVVG (69\%), SVJ (65\%) and SVLS (61\%). \textit{IV}, \textit{C\_PV}, \textit{Gamma} and \textit{Theta} are significant in forecasting the PER of all six models. The PER of all models is negatively related to the previous day's \textit{IV}, \textit{C\_PV}, \textit{Gamma} and \textit{Theta}. PER is also inversely linked to \textit{DDelta} and \textit{DSPXV}. The estimated coefficients of \textit{IV}, \textit{C\_PV}, ``greeks'' (\textit{DDelta}, \textit{Gamma} and \textit{Theta}) and \textit{DSPXV} indicate that the implied volatility level, the difference between call and put options trades, the sensitivity of the option prices and the number of the underlying trades are negatively linked to the PER of models. \textit{C\_PP} has a significant effect on the PER of stochastic volatility models while \textit{H\_L} reveals a significant effect on the PER of SV, SVJ, SVCJ and SVLS.
 \begin{table}[!ht]
    \centering
    \scriptsize
    \caption{Forecasting Model Risk}\label{fore_mr}
    \setlength{\tabcolsep}{2mm}{
   \begin{tabular}{@{}lllllllllll@{}}
\toprule
    & C             & IV            & C\_PV           & DDelta           & Gamma           & Theta          & H\_L            & DSPXV          & C\_PP           & Adj. $R^2$     \\ \midrule
\multicolumn{11}{l}{Panel A. MJD}                                                                                                                                             \\ \midrule
PER  &  {2.42**}   &  {-5.70**}  &  {-0.09**} & -3.14            &  {-83.63**}  &  {-0.30**} & 0.05                    &  {-0.06**} & 0.25            & 0.80      \\
MSR  &  {-12.67**} &  {51.51**}  &  {-0.75**} &  {-258.87**} &  {298.83**}  &  {-3.98**} &  {4.01**}           &  {-0.61**} &  {-23.04**} & 0.62      \\
L\_S &  {0.60**}   &  {-2.57**}  &  {-0.06**} &  {-8.36**}   &  {-39.33**}  &  {-0.39**} &  {0.24**}           &  {-0.06**} &  {-0.57**}  & 0.84      \\ \midrule
\multicolumn{11}{l}{Panel B. SV}                                                                                                                                                          \\ \midrule
PER  &  {3.74**}   &  {-6.65**}  &  {-0.08**} &  {-118.52**} &  {-158.29**} &  {-1.07**} &  {-0.24*}          &  {-0.32**} &  {-1.45**}  & 0.58      \\
MSR  & 0.63            &  {23.97**}  &  {0.07**}  & -60.21           &  {-229.39**} &  {-1.29**} & 0.08                    &  {-0.72**} & -3.04           & 0.46      \\
L\_S &  {-18.80**} &  {67.22**}  &  {0.59**}  & -166.43          &  {1088.37**} &  {7.67**}  & 0.70                    &  {3.24*}  &  {48.65**}  & 0.08      \\ \midrule
\multicolumn{11}{l}{Panel C. SVJ}                                                                                                                                                         \\ \midrule
PER  &  {8.79**}   &  {-20.28**} &  {-0.33**} &  {-65.80**}  &  {-294.09**} &  {-0.95**} &  {-0.59**} &  {-0.37**} &  {2.76**}   & 0.65      \\
MSR  &  {-5.27**}  &  {24.82**}  &  {-0.26**} & -9.66            &  {159.58**}  &  {-0.25*} &  {1.68**}           &  {-0.28*} &  {-7.08**}  & 0.47      \\
L\_S &  {14.08**}  &  {-14.29**} & -0.20          &  {-290.64*} &  {-529.40**} &  {3.45**}  &  {-6.47**}          & -0.61          &  {15.66*}  & 0.06      \\ \midrule
\multicolumn{11}{l}{Panel D. SVCJ}                                                                                                                                                        \\ \midrule
PER  &  {7.37**}   &  {-16.95**} &  {-0.29**} &  {-54.65**}  &  {-249.62**} &  {-0.95**} &  {0.73**}           & -0.04          &  {2.96**}   & 0.56      \\
MSR  &  {-4.36**}  &  {23.33**}  &  {-0.31**} & -23.31           &  {110.74**}  &  {-0.41**} & -0.55                   &  {-0.41*} &  {-8.22**}  & 0.38      \\
L\_S &  {22.34**}  &  {-43.31**} &  {-1.13**} & 35.71            &  {-849.71**} &  {3.31**}  &  {-19.05**}         &  {-3.55**} &  {-46.89**} & 0.08      \\ \midrule
\multicolumn{11}{l}{Panel E. SVVG}                                                                                                                                                        \\ \midrule
PER  &  {8.42**}   &  {-18.72**} &  {-0.32**} &  {-33.11*}  &  {-277.10**} &  {-0.63**} & -0.15                   &  {-0.28**} &  {3.19**}   & 0.69      \\
MSR  &  {4.11**}   &  {-12.34**} &  {-1.13**} & -30.67           & 47.03            & -0.02          & 1.21                    &  {-0.62*} & 4.03            & 0.14      \\
L\_S &  {-47.53**} &  {100.65**} &  {1.85**}  & 106.19           &  {2527.45**} &  {2.43**}  &  {-6.83**}          &  {1.82*}  &  {-27.19**} & 0.32      \\ \midrule
\multicolumn{11}{l}{Panel F. SVLS}                                                                                                                                                        \\ \midrule
PER  &  {2.52**}   &  {-6.06**}  &  {-0.07**} &  {-64.86**}  &  {-108.56**} &  {-1.18**} &  {0.31**}           &  {-0.20**} &  {-4.24**}  & 0.61      \\
MSR  &  {-4.51**}  &  {35.09**}  &  {-0.12**} &  {-123.56**} & 5.41             &  {-0.79**} & -0.47                   &  {-0.67**} &  {-4.15*}  & 0.56      \\
L\_S &  {14.41**}  &  {19.79**}  &  {0.65**}  & -27.19           & -215.56          &  {14.38**} &  {-9.22**}          & 2.06           &  {82.10**}  & 0.17      \\ \bottomrule
\end{tabular}}
    \begin{tablenotes}
      \scriptsize
      \item NOTE: The regression results are based on Equation (\ref{reg_fore}). * and ** indicate values significant at 5\% and 1\% significance levels, respectively.
    \end{tablenotes}
    \end{table}

Compared with PER, the MSR is more difficult to predict. The adjusted $R^2$ is higher for the MSR of MJD and SVLS (62\% and 56\% respectively), and lower for SV and SVJ at just above 46\%; it drops remarkably for SVCJ (38\%) and it is only 14\% for SVVG. \textit{IV}, \textit{C\_PV} and \textit{DSPXV} are significant in predicting MSR of all models. MSR tends to increase with \textit{IV}, except for SVVG, whose MSR decreases with \textit{IV}. Moreover, except for SV, the MSR of all other five models shows a significantly negative relationship with \textit{C\_PV}. The MSR of all models decreases with \textit{DSPXV}, indicating that an increase in the number of underlying trades decreases the MSR. Option ``greeks'' have significant power in forecasting the MSR of MJD.

Most of the L\_S of MJD can be predicted with these factors with an adjusted $R^2$ value of 84\%, followed by SVVG (32\%) and SVLS (17\%). The values of adjusted $R^2$ for L\_S are only around 8\% for SV and AJD models. \textit{IV}, \textit{Theta} and \textit{C\_PP} have significant power in forecasting the L\_S for all six models. \textit{IV}, \textit{C\_PV} and \textit{DSPXV} are negatively linked to the L\_S of MJD and AJD models and positively linked to the L\_S of SV, SVVG and SVLS. The coefficients of \textit{Theta} for the L\_S of stochastic volatility models are all positive. 
Moreover, the L\_S of SV, SVJ and SVLS increase with \textit{C\_PP} while the L\_S of MJD, SVCJ and SVVG decrease with \textit{C\_PP}.

Overall, regardless of the model, \textit{IV}, \textit{C\_PV}, \textit{Gamma}, \textit{Theta} have statistically significant power in forecasting PER; \textit{IV}, \textit{C\_PV} and \textit{DSPXV} are powerful in predicting MSR; besides, \textit{IV}, \textit{Theta} and \textit{C\_PP} can help in predicting L\_S. \textit{IV} is the most effective factor in forecasting model risk, coming out as significant in all regressions, while \textit{C\_PV} and \textit{Theta} are also useful in forecasting model risk.

\section{Conclusions}

In this paper, we propose an ES-type model risk measure which is able to estimate PER and MSR of continuous-time pricing models. We then apply this measurement to L\'{e}vy jump models and AJD models to investigate to what extent the MSR and PER affect the models' ability to capture the joint dynamics of stock and option prices. Building on the approaches of \cite{broadie2007model} and \cite{yu2011mcmc}, we develop an effective MCMC method to jointly estimate parameters and latent variables with both stock and option prices. 

We show that the PER is dominant for stochastic volatility models. The introduction of jumps increases PER and decreases MSR. We find that SVLS has the smallest TMR. SVVG has less MSR and PER when the market is turbulent, but its model risk soars when the market is calm. By contrast, AJD models and SVLS tend to have slightly higher model risk than SVVG under volatile periods while they carry small MSR in tranquil periods. We also find that short positions tend to have a higher model risk when using stochastic volatility models for call options when the market is turbulent. We further show that AJD models have significantly lower MSR than L\'{e}vy jump models, but the option pricing errors of L\'{e}vy jump models and AJD models reveal few significant differences. 

We find that market risk factors can predict more than half of the PER of models. PER decreases with the previous day's implied volatility, underlying trading volume and option greeks. The MSR of AJD models and L\'{e}vy jump models is difficult to predict. The previous day's implied volatility, the difference between trading volumes of call and put options and the theta sensitivity of options are the most decisive factors in forecasting model risk.

Our results highlight that it is necessary to measure PER and MSR separately, as the two
components of TMR. A model with small MSR is not necessarily the most desirable if the parameters are likely to be very difficult to estimate due to the complexity of the model. Similarly, a model with low PER tends to be simple but it is likely to have large MSR. The best models have a tradeoff between PER and MSR. A distinction between PER and MSR can enhance model risk management. In addition, model risk can influence model parameter values, and this dependence can be explicitly modeled, as in \cite{bollerslev2016exploiting}; in future research, it would be interesting to explore the dependence of parameter values on PER and MSR separately. Furthermore, non-parametric models can learn from the model risk of parametric models, and their pricing performance can be improved as suggested by \cite{fan2009option}. It remains an open question how the pricing performance can be improved using information on MSR and PER.

\newpage
 \begin{spacing}{1.5}
\bibliographystyle{jfe}
\bibliography{ref_paper}
 \end{spacing}
\newpage
\begin{appendices}

\newcommand{\itemEq}[1]{%
         \begingroup%
         \setlength{\abovedisplayskip}{0pt}%
         \setlength{\belowdisplayskip}{0pt}%
         \parbox[c]{\linewidth}{\begin{flalign}#1&&\end{flalign}}%
         \endgroup}

\setcounter{table}{0}
\renewcommand{\thetable}{A\arabic{table}}

\setcounter{figure}{0}
\renewcommand{\thefigure}{A\arabic{figure}}

\setcounter{equation}{0}
\renewcommand{\theequation}{A\arabic{equation}}

\section{The Jump Characteristic Component and Priors for Model Parameters}\label{appA}

The jump component $\Phi_J(u)$ in the characteristic function of different models is provided below.

SVCJ: $\Phi_J(u) = \lambda \left( 1-\frac{\exp\left(iu\mu_J^\mathbb{Q} - \frac{1}{2}\sigma_J^2 u^2 \right)}{1 - i u \mu_V \rho_J - iu\mu_V} \right)$.

The expression of $\Phi_J(u)$ for SVJ can be derived by taking $\mu_V=0$ in the equation above.

SVVG: $\Phi_J(u) = \frac{\ln \left( 1 - iu\gamma^\mathbb{Q} \nu + \frac{(\sigma^\mathbb{Q})^2 \nu u^2}{2}  \right) }{\nu}$.

SVLS: $\Phi_J(u) = (\sigma |u|)^\alpha \left(1+i \times {\rm sign}(u) \times \tan \left(\frac{\pi \alpha}{2}\right)\right)$, where ${\rm sign}(u)$ extracts the sign of $u$.

 The priors of parameters for all models are detailed here. We choose uninformative and conditionally conjugate priors whenever possible. Priors for common parameters: $\kappa \sim \mathbb{N}(0, 1)\textbf{1}_{\kappa > 0}$, $\theta \sim \mathbb{N} (0, 1) \textbf{1}_{\theta > 0}$, $\rho \sim \mathbb{U}(-1, 1)$, $\sigma_v^2 \sim \mathbb{IG}(2.5, 0.1)$, $\eta_s \sim \mathbb{N}(0, 100)$, $\eta_v \sim \mathbb{N}(0, 100)$, $\rho_c \sim \mathbb{N}(0, 1)$, $\sigma_c^2 \sim \mathbb{IG}(2.5, 0.1)$. Priors for parameters common to SVJ and SVCJ: $\mu_J^\mathbb{P} \sim \mathbb{N}(0, 100)$, $\mu_J^\mathbb{Q} \sim \mathbb{N}(0, 100)$, $\sigma_J^2 \sim \mathbb{IG}(10, 40)$ and $\lambda \sim \mathbb{B}(2, 40)$. Priors for parameters unique to SVCJ: $\mu_V \sim \mathbb{IG}(10, 20)$ and $\rho_J \sim \mathbb{N}(0, 4)$. Priors for parameters unique to SVVG: $\gamma^\mathbb{P} \sim \mathbb{N}(0, 1)$, $\gamma^\mathbb{Q} \sim \mathbb{N}(0, 1)$, $\nu \sim \mathbb{IG}(10, 20)$, $(\sigma^\mathbb{P})^2 \sim \mathbb{IG}(2.5, 0.1)$ and $(\sigma^\mathbb{Q})^2 \sim \mathbb{IG}(2.5, 0.1)$. Priors for parameters unique to SVLS: $\alpha \sim \mathbb{U}(1,2)$ and $\sigma^{\frac{\alpha}{\alpha-1}}|\alpha \sim \mathbb{IG}(2.5, 0.1)$, where $\mathbb{N}$ refers to the Normal distribution, $\mathbb{IG}$ refers to the Inverse Gamma distribution, $\mathbb{U}$ refers to a standard Uniform distribution, and $\mathbb{B}$ to a Beta distribution. These priors have similar values to those used in \cite{eraker2003impact}, \cite{li2006bayesian} and \cite{yu2011mcmc}. We also investigate various other priors, and the result does not change significantly.

\section{MCMC methods for Parameter Estimation} \label{appB}
This section introduces the updating algorithms and posterior distributions of model parameters and latent variables. 

We focus on SVCJ at first, the estimation of SVJ can be derived by setting $\mu_V=0$ while taking $\lambda=0$ leads to the estimation of SV. MCMC methods for estimating unique parameters and latent variables in SVVG and SVLS are also introduced in this section.

Posterior for $\kappa$: $\kappa| {Y, V, J^Y, \Theta{\backslash\{\kappa\}}} \sim \mathbb{N}\left(\frac{\mathcal{A}}{\mathcal{B}}, \sqrt{\frac{1}{\mathcal{B}}}\right)\mathbf{1}_{\kappa>0}$, where  $\mathcal{B} = \frac{\Delta}{(1-\rho^2)\sigma_v^2} \sum_{t=0}^{T-1} \frac{(\theta - V_t)^2}{V_t} + 1$; $\mathcal{A} = \frac{1}{\sigma_v (1 - \rho^2)} \sum_{t=0}^{T-1} \frac{(\theta - V_t)\frac{V_{t+1} - V_t}{\sigma_v} - \rho H_{t+1}}{V_t}$; and $H_{t+1} = Y_{t+1} - Y_t - \left(r_t - \frac{1}{2} V_t + \Phi_J(-i) + \eta_s V_t\right)\Delta - J_{t+1}^Y$.

Posterior for $\theta$: $\theta| {Y, V, J^Y, J^V, \Theta{\backslash\{\theta\}}} \sim \mathbb{N}\left(\frac{\mathcal{A}}{\mathcal{B}}, \sqrt{\frac{1}{\mathcal{B}}}\right)\mathbf{1}_{\theta>0}$,
  where $\mathcal{B} = \frac{\kappa^2 \Delta}{(1-\rho^2)\sigma_v^2} \sum_{t=0}^{T-1} \frac{1}{V_t} + 1$; $\mathcal{A} = \frac{\kappa}{\sigma_v (1 - \rho^2)} \sum_{t=0}^{T-1} \frac{\frac{ D_{t+1}} {\sigma_v} - \rho H_{t+1}}{V_t}$; $D_{t+1} = V_{t+1} + (\kappa \Delta -1)V_t - J_{t+1}^V$; and $H_{t+1} = Y_{t+1} - Y_t - (r_t - \frac{1}{2} V_t + \Phi_J(-i) + \eta_s V_t)\Delta - J_{t+1}^Y$.

 Posterior for $\sigma_v^2$: The conditional posteriors of $\sigma_v^2$ cannot be written as a standard distribution; to overcome this difficulty, we apply an independent Metropolis-Hasting algorithm. The conditional posterior of $\sigma_v^2$, conditional on $Y$ $V$, $J^Y$, $J^V$, and $\Theta{\backslash\{\sigma_v\}}$, is proportional to $(\sigma_v^2)^{\frac{T + c^* +2}{2}} \times \pi (\sigma_v) \times \exp \left \{ -\frac{1}{2} \frac{C^*}{\sigma_v^2} \right \}$, and $\pi (\sigma_v) := \exp \left \{ -\frac{1}{2} \sum_{t=0}^{T-1} \frac{\left( D_{t+1} - \rho \sigma_v (H_{t+1})\right)^2 } {(1-\rho^2) \sigma_v^2 V_t \Delta} \right \}$,
  where $c^*=2.5$, and $C^*=0.1$ are hyperparameters of the prior of $\sigma_v^2$; $D_{t+1} = V_{t+1} - V_t - \kappa (\theta - V_t)\Delta - J_{t+1}^V$ and $H_{t+1} = Y_{t+1} - Y_t - \left(r_t - \frac{1}{2} V_t + \Phi_J(-i) + \eta_s V_t\right)\Delta - J_{t+1}^Y$. Although the posterior distribution is non-standard, it becomes a one-dimensional inverted-Gamma distribution if there is no leverage effect ($\rho = 0$). Therefore, we set the proposal density as: $\sigma_v^2| {Y, V, J^V, \Theta{\backslash\{\sigma_v\}}} \sim \mathbb{IG}(\mathcal{A},\mathcal{B})$,
  where $\mathcal{A} = c^*+T$;
  $\mathcal{B} = C^* + \sum_{t=0}^{T-1} \frac{D_{t+1}^2}{V_t \Delta}$;
  and $D_{t+1} = V_{t+1} - V_t - \kappa (\theta - V_t) \Delta - J_{t+1}^V$.
  For a given previous draw $\sigma_v^{(g)}$, we draw $\sigma_v^{(g+1)}$ from the proposal density function, and then accept $\sigma_v^{(g+1)}$ with probability $\min \left ( \frac{\pi (\sigma_v^{(g+1)})}{\pi(\sigma_v^{(g)})}, 1  \right)$.

 Posterior for $\rho$ $\propto \pi (\rho) := (1-\rho^2) ^ {-\frac{T}{2}} \exp \{ -12 (1-\rho^2) \sum_{t=0}^{T-1} (H_{t+1}^2 + D_{t+1}^2) + \frac{\rho}{1-\rho^2} \sum_{t=0}^{T-1} \\ \ H_{t+1} D_{t+1}\}$,
  where $D_{t+1} = \frac{V_{t+1} - V_t - \kappa(\theta -V_t) \Delta - J_{t+1}^V}{\sigma_v \sqrt{V_t \Delta}}$ and $H_{t+1} = \frac{Y_{t+1} - Y_t - \left(r_t - \frac{1}{2} V_t + \Phi_J(-i) + \eta_s V_t\right)\Delta - J_{t+1}^Y}{\sqrt{V_t \Delta}}$. The estimation is updated using following algorithm:
  (1) Draw $\frac{1}{2} \ln \frac{1+ \rho ^{(g+1)}}{1-\rho^{(g+1)}}$ from $\mathbb{N}\left(\frac{1}{2} \ln \frac{1+\rho_r}{1-\rho_r}, \frac{1}{T-3}\right)$, where $\rho_r = Corr (\mathbf{D}, \mathbf{H})$, where $\mathbf{D} = \{D_{t+1}\}_{t=0}^{T-1}$, $\mathbf{H} = \{H_{t+1}\}_{t=0}^{T-1}$, and $Corr$ denotes correlation;
  (2) accept $\rho^{g+1}$ with $\min \left ( \frac{\pi \left(\rho ^{(g+1)}\right)}{\pi \left(\rho ^{(g)}\right)}  \times \frac{\exp \left( -\frac{\left(f\left(\rho^{(g)}\right) - f(\rho_r)\right)^2}{\frac{2}{T-3}} \right)}{\exp\left( -\frac{\left(f\left(\rho^{(g+1)}\right) - f(\rho_r)\right)^2}{\frac{2}{T-3}} \right)}, 1  \right)$, where $f(\rho) = \frac{1}{2} \ln \frac{1+\rho}{1-\rho}$.

Posterior for $\mu_J^\mathbb{P}$: $\mu_J^\mathbb{P}| {\upxi^Y,\upxi^V, \Theta{\backslash\{\mu_J^\mathbb{P}\}}} \sim \mathbb{N}\left(\frac{\mathcal{A}}{\mathcal{B}}, \sqrt{\frac{1}{\mathcal{B}}}\right)$,
  where $\mathcal{A} =  \sum_{t=0}^{T-1} \frac{\left(\upxi_{t+1}^Y - \rho_J \upxi_{t+1}^V\right)}{\sigma_J^2} + \frac{c^*}{C^*}$;
  $\mathcal{B} =\frac{T}{\sigma_J^2} + \frac{1}{C^*} $;
  $c^*=0$ and $C^*=100$ are hyperparameters of the prior of $\mu_J^\mathbb{P}$.

Posterior for $\sigma_J^2$: $\sigma_J^2| {\upxi^Y,\upxi^V, \Theta{\backslash\{\sigma_J\}}} \sim \mathbb{IG}(\mathcal{A},\mathcal{B})$,
  where $\mathcal{A} = c^* + T$;
  $\mathcal{B} =C^* + \sum_{t=0}^{T-1} (\upxi_{t+1}^Y -  \mu_J^\mathbb{P} - \rho_J \upxi_{t+1}^V)^2$;
  $c^*=10$ and $C^*=40$ are hyperparameters of the prior of $\sigma_J^2$.

Posterior for $\lambda$: $\lambda| {N} \sim \mathbb{B}(\mathcal{A},\mathcal{B})$,
  where $\mathcal{A} = c^* + \sum_{t=0}^{T-1} N_{t+1}$;
  $\mathcal{B} =C^* + T - \sum_{t=0}^{T-1} N_{t+1}$;
  $c^*=2$ and $C^*=40$ are hyperparameters of the prior of $\lambda$.

  Posterior for $N$: $N_{t+1}| {Y, V, \upxi^Y,\upxi^V, \Theta} \sim Bernoulli \left(\frac{\alpha_1}{\alpha_1 + \alpha_2}\right)$. $Bernoulli$ denotes a Bernoulli distribution. $D_{t+1} = \frac{V_{t+1} - V_t - \kappa(\theta -V_t) \Delta - \upxi_{t+1}^V}{\sigma_v \sqrt{V_t \Delta}}$; $H_{t+1} = \frac{Y_{t+1} - Y_t - \left(r_t - \frac{1}{2} V_t + \Phi_J(-i) + \eta_s V_t\right)\Delta - \upxi_{t+1}^Y}{\sqrt{V_t \Delta}}$;
   $\alpha_1 = \lambda \exp\left( -\frac{1}{2 (1-\rho^2)} \left(H_{t+1}^2  - 2 \rho H_{t+1} D_{t+1}\right) \right)$; $HH_{t+1} = \frac{Y_{t+1} - Y_t - \left(r_t - \frac{1}{2} V_t + \Phi_J(-i) + \eta_s V_t\right)\Delta}{\sqrt{V_t \Delta}}$; $DD_{t+1} = \frac{V_{t+1} - V_t - \kappa(\theta -V_t) \Delta }{\sigma_v \sqrt{V_t \Delta}}$;  and
  $\alpha_2 = (1-\lambda) \exp\left( -\frac{1}{2 (1-\rho^2)} \left(HH_{t+1}^2 - 2 \rho HH_{t+1} DD_{t+1}\right) \right)$.

 Posterior for $\eta_v$ $\propto  \pi (\eta_v) := \prod_{t=0}^{T-1} \exp \left( - \frac{\left[(C_{t+1} - F_{t+1}) - \rho_c (C_t - F_t)\right]^2}{2\sigma_c^2} \right) \times \exp \left( -\frac{\eta_v^2}{2} \right)$.

 Posterior for $\mu_J^\mathbb{Q}$ $\propto  \pi \left(\mu_J^\mathbb{Q}\right) := \prod_{t=0}^{T-1} \exp \left( - \frac{\left[(C_{t+1} - F_{t+1}) - \rho_c (C_t - F_t)\right]^2}{2\sigma_c^2} \right) \times \exp \left( -\frac{\left(\mu_J^\mathbb{Q}\right)^2}{2} \right)$,
  $\eta_v$ and $\mu_J^\mathbb{Q}$ are estimated with the Metropolis-Hasting algorithm. 

 Posterior for $\eta_s$: $\eta_s|{Y, V, J^Y, J^V, \Theta{\backslash\{\eta_s\}}} \sim \mathbb{N}\left(\frac{\mathcal{A}}{\mathcal{B}}, \sqrt{\frac{1}{\mathcal{B}}}\right)$,
where $\mathcal{B} = \frac{\Delta}{(1-\rho^2)} \sum_{t=0}^{T-1} V_t + 1$; $\mathcal{A} = \frac{1}{(1 - \rho^2)} \sum_{t=0}^{T-1} \left(H_{t+1} - \frac{\rho}{\sigma_v} D_{t+1}\right)$; $D_{t+1} = V_{t+1} - V_t - \kappa (\theta - V_t)\Delta - J_{t+1}^V$;
  and $H_{t+1} = Y_{t+1} - Y_t - (r_t - \frac{1}{2} V_t + \Phi_J(-i) + \eta_s V_t)\Delta - J_{t+1}^Y$.

 Posterior for $\mu_V$: $\mu_V|{\upxi^V} \sim \mathbb{IG}(\mathcal{A}, \mathcal{B})$,
where $\mathcal{A} = c^* + 2T$;
  $\mathcal{B} = C^* + 2 \sum_{t=0}^{T-1} \upxi_{t+1}^V$;
  $c^*=10$ and $C^*=20$ are hyperparameters of the prior of $\mu_V$.

Posterior for $\rho_J$: $\rho_J|{ \upxi^Y, \upxi^V, \Theta{\backslash\{\rho_J\}}} \sim \mathbb{N}\left(\frac{\mathcal{A}}{\mathcal{B}}, \sqrt{\frac{1}{\mathcal{B}}}\right)$,
where $\mathcal{A} =  \frac{\sum_{t=0}^{T-1} \left(\upxi_{t+1}^Y - \mu_J^\mathbb{P}\right)\upxi_{t+1}^V}{\sigma_J^2} + \frac{c^*}{C^*} $;
  $\mathcal{B} = \frac{\sum_{t=0}^{T-1} \left(\upxi_{t+1}^V\right)^2}{\sigma_J^2} + \frac{1}{C^*}$;
  $c^*=0$ and $C^*=4$ are hyperparameters of the prior of $\rho_J$.

  Posterior for $\rho_c$: $\rho_c|{ C, F, \Theta{\backslash\{\rho_c\}}} \sim \mathbb{N}\left(\frac{\mathcal{A}}{\mathcal{B}}, \sqrt{\frac{1}{\mathcal{B}}}\right)$,
where $\mathcal{A} =  \frac{\sum_{t=0}^{T-1} H_t H_{t+1}}{\sigma_c^2} + \frac{c^*}{C^*} $;
  $\mathcal{B} = \frac{\sum_{t=0}^{T-1}H_t^2}{\sigma_c^2} + \frac{1}{C^*}$;
  $H_{t+1} = C_{t+1} - F_{t+1}$;
  $c^*=0$ and $C^*=1$ are hyperparameters of the prior of $\rho_c$.

  Posterior for $\sigma_c$: $\sigma_c^2|{ C, F, \Theta{\backslash\{\sigma_c\}}} \sim \mathbb{IG}(\mathcal{A},\mathcal{B})$,
where $\mathcal{A} =  c^* + T $;
  $\mathcal{B} = C^* + \sum_{t=0}^{T-1} (H_{t+1} - \rho_c H_t)^2$;
   $H_{t+1} = C_{t+1} - F_{t+1}$;
  $c^*=2.5$ and $C^*=0.1$ are hyperparameters of the prior of $\sigma_c^2$.

  Posterior for $\upxi^Y$: $\upxi_{t+1}^Y| {Y, V, \upxi^V, N_{t+1}=1, \Theta} \sim \mathbb{N}\left(\frac{\mathcal{A}}{\mathcal{B}}, \sqrt{\frac{1}{\mathcal{B}}}\right)$,
  where $\mathcal{A} = \frac{H_{t+1} - \frac{\rho}{\sigma_v} D_{t+1}}{(1-\rho^2)V_t \Delta} + \frac{\mu_J^\mathbb{P} + \rho_J D_t+1}{\sigma_J^2}$;
  $\mathcal{B} = \frac{1}{(1-\rho^2) V_t \Delta} + \frac{1}{\sigma_J^2} $;
  $D_{t+1} = V_{t+1} - V_t - \kappa (\theta - V_t)\Delta - \upxi_{t+1}^V$;
  and $H_{t+1} = Y_{t+1} - Y_t - (r_t - \frac{1}{2} V_t + \Phi_J(-i) + \eta_s V_t)\Delta $.

   Posterior for $\upxi^V$: $\upxi_{t+1}^V|{Y, V, \upxi^Y, N_{t+1} = 1, \Theta} \sim \mathbb{N}\left(\frac{\mathcal{A}}{\mathcal{B}}, \sqrt{\frac{1}{\mathcal{B}}}\right)\mathbf{1}_{\upxi_{t+1}^V>0}$,
  where $\mathcal{A} = \frac{D_{t+1} - \rho \sigma_v H_{t+1}}{(1-\rho^2) \sigma_v^2 V_t \Delta} + \frac{\rho_J \left(H_{t+1} - \mu_J^\mathbb{P}\right)}{\sigma_J^2} - \frac{1}{\mu_V}$;
  $\mathcal{B} = \frac{1}{(1-\rho^2) \sigma_v^2 V_t \Delta} + \frac{\rho_J^2}{\sigma_J^2} $;
  $D_{t+1} = V_{t+1} - V_t - \kappa (\theta - V_t)\Delta$;
  and $H_{t+1} = Y_{t+1} - Y_t - (r_t - \frac{1}{2} V_t + \Phi_J(-i) + \eta_s V_t)\Delta - \upxi_{t+1}^Y $.

   Posterior for $V$ $\propto \pi (V_{t+1}) := \frac{1}{V_{t+1}} \times \exp \left\{ -\frac{-2\rho D_{t+1} H_{t+1} + D_{t+1}^2}{2(1-\rho^2)} - \frac{D_{t+2}^2 - 2\rho D_{t+2} H_{t+2} +H_{t+2}^2 }{2(1-\rho^2)} \right)$, for $0<t+1<T$.
   $D_{t+1} = \frac{V_{t+1} - V_t - \kappa (\theta - V_t)\Delta - J_{t+1}^V}{\sigma_v \sqrt{V_t \Delta}}$;
  and $H_{t+1} = \frac{Y_{t+1} - Y_t - \left(r_t - \frac{1}{2} V_t + \Phi_J(-i) + \eta_s V_t\right)\Delta - J_{t+1}^Y}{\sqrt{V_t \Delta}}$.

  $V_0$ and $V_T$ are estimated in a similar way. This is a nonstandard distribution, and we apply the random-walk Metropolis-Hastings algorithm with the Student's $t$ distribution, the average standard deviations is 0.25, the degrees of freedom is 6.

Below we describe the algorithm to estimate parameters and latent variables unique to SVVG.

Posterior for $\gamma^\mathbb{P}$: $\gamma^\mathbb{P}|{ G, J^Y, \Theta{\backslash\{\gamma^\mathbb{P}\}}} \sim \mathbb{N}\left(\frac{\mathcal{A}}{\mathcal{B}}, \sqrt{\frac{1}{\mathcal{B}}}\right)$,
where $\mathcal{A} =  \frac{1}{\left(\sigma^\mathbb{P}\right)^2} J_{t+1}^Y + \frac{c^*}{C^*}$;
  $\mathcal{B} = \frac{1}{\sigma^\mathbb{P})^2} G_{t+1} \frac{1}{C^*}$;
  $c^*=0$ and $C^*=1$ are hyperparameters of the prior of $\gamma^\mathbb{P}$.

Posterior for $\sigma^\mathbb{P}$: $(\sigma^\mathbb{P})^2|{ G, J^Y, \Theta{\backslash\{\sigma^\mathbb{P}\}}} \sim \mathbb{IG}(\mathcal{A},\mathcal{B})$,
where $\mathcal{A} =  c^* +T$;
  $\mathcal{B} = C^* + \sum_{t=0}^{T-1} \frac{(J_{t+1}^Y - \gamma^\mathbb{P} G_{t+1})^2}{G_{t+1}}$;
  $c^*=2.5$ and $C^*=0.1$ are hyperparameters of the prior of $\left(\sigma^\mathbb{P}\right)^2$.

  Posterior for $\nu$ $\propto \pi (\nu) := \left( \frac{1}{\nu ^{\frac{\Delta}{\nu}} \Gamma(\frac{\Delta}{\nu})} \right)^T \left( \prod_{t=0}^{T-1} G_{t+1} \right)^{\frac{\Delta}{\nu}} \exp \left\{ - \frac{1}{\nu} \left( \sum_{t=0}^{T-1} G_{t+1} + C^* \right) \right\} \left( \frac{1}{\nu} \right)^{c^* +1}$,
  $c^*=10$ and $C^*=20$ are hyperparameters of the prior of $\nu$. $\Gamma$ denotes the Gamma function.

   Posterior for $\gamma^\mathbb{Q}$ $\propto  \pi (\gamma^\mathbb{Q}) := \prod_{t=0}^{T-1} \exp \left( - \frac{[(C_{t+1} - F_{t+1}) - \rho_c (C_t - F_t)]^2}{2\sigma_c^2} \right) \times \exp \left( -\frac{\left(\gamma^\mathbb{Q}\right)^2}{2} \right)$.

    Posterior for $\sigma^\mathbb{Q}$ $\propto  \pi (\sigma^\mathbb{Q}|{ \sigma^\mathbb{Q}>0}) := \prod_{t=0}^{T-1} \exp \left( - \frac{[(C_{t+1} - F_{t+1}) - \rho_c (C_t - F_t)]^2}{2\sigma_c^2} \right) \times \exp \left( -\frac{\left(\sigma^\mathbb{Q}\right)^2}{2} \right)$,
   $\nu$, $\gamma^\mathbb{Q}$ and $\sigma^\mathbb{Q}$ are estimated with the Metropolis-Hasting algorithm.

  Posterior for $J^Y$: $J_{t+1}^Y| {Y, V, G, \Theta} \sim \mathbb{N}\left(\frac{\mathcal{A}}{\mathcal{B}}, \sqrt{\frac{1}{\mathcal{B}}}\right)$,
  where $\mathcal{A} = \frac{1}{(1-\rho^2) V_t \Delta} \left( H_{t+1} -\frac{\rho D_{t+1}}{\sigma_v} \right)+ \frac{\gamma^\mathbb{P}}{(\sigma^\mathbb{P})^2} $;
  $\mathcal{B} = \frac{1}{(1-\rho^2) V_t \Delta} + \frac{1}{(\sigma^\mathbb{P})^2G_{t+1}} $;
  $D_{t+1} = V_{t+1} - V_t - \kappa (\theta - V_t)\Delta $;
  and $H_{t+1} = Y_{t+1} - Y_t - (r_t - \frac{1}{2} V_t + \Phi_J(-i) + \eta_s V_t)\Delta $.

    Posterior for $G$ $\propto \pi (G_{t+1}):= G_{t+1}^{\frac{\Delta}{\nu} - \frac{3}{2}} \exp\left\{ -\frac{J_t^2}{2(\sigma^\mathbb{P})^2 G_{t+1}} - G_{t+1} \left( \frac{\left(\gamma^\mathbb{P}\right)^2}{2\left(\sigma^\mathbb{P}\right)^2} + \frac{1}{\nu} \right) \right\}$.
We sample $G_{t+1}$ as $G_{t+1} |{J^Y, \Theta} \sim \mathbb{GIG}\left(\frac{1}{\nu}-\frac{1}{2}, \frac{(\gamma^\mathbb{P})^2}{(\sigma^\mathbb{P})^2} + \frac{2}{\nu}, \frac{(J_{t+1}^Y)^2}{(\sigma^\mathbb{P})^2}\right)$ directly, where $\mathbb{GIG}$ denotes a generalized inverse Gaussian distribution.

The posterior distribution of the parameters and latent variables unique to SVLS is calculated as follows.

Posterior for $\alpha$ $\propto \pi (\alpha) :=   \left( \frac{\alpha}{\alpha-1} \right)^T \exp\left\{ - \sum_{t=0}^{T-1} \left| \frac{J_{t+1}^Y}{\sigma \Delta ^ \frac{1}{\alpha} t_\alpha (U_{t+1})} \right|^{\frac{\alpha}{\alpha-1}}  \right\} \times \prod_{t=0}^{T-1} \left|  \frac{J_{t+1}^Y}{\sigma \Delta ^ \frac{1}{\alpha} t_\alpha (U_{t+1})} \right|^{\frac{\alpha}{\alpha-1}} \times \left[ (\frac{1}{\sigma})^{\frac{\alpha}{\alpha-1}} \right]^{c^*+1} \times \exp\left\{ -(\frac{1}{\sigma})^{\frac{\alpha}{\alpha-1}} C^* \right\} \times \mathbf{1}_{1.01\leq \alpha \leq 2}$,
where $c^* = 2.5$ and $C^*=0.1$ are hyperparameters of the prior of $\sigma^{\frac{\alpha}{\alpha-1}}$, and $t_\alpha (U_{t+1}) = \left( \frac{\sin \left[\pi \alpha U_{t+1} + \frac{(2-\alpha)\pi}{2} \right]}{cos[\pi U_{t+1}]} \right) \times \left( \frac{cos[\pi U_{t+1}]}{cos\left[ \pi (\alpha -1)U_{t+1} + \frac{(2-\alpha)\pi}{2} \right]} \right)^{\frac{\alpha-1}{\alpha}}$. The posterior distribution of $\alpha$ is nonstandard, we simulate the posterior based on the Metropolis-Hastings method and take a linearly transformed Beta distribution as the proposal density:
  (1) draw $\varpi$ from $\mathbb{B}(\mathcal{A}, \mathcal{B})$, where $\mathcal{A} = \frac{\alpha^{(g)}-1.01}{0.99} (5 \ln(T) -2) +1$ and $\mathcal{B} = 5 \ln(T) - \mathcal{A}$. Set $\alpha^{(g+1)} = 0.99\varpi + 1.01$;
  (2) calculate $\mathcal{A}^* = \frac{\alpha^{(g+1)}-1.01}{0.99} (5 \ln(T) -2) +1$ and $\mathcal{B}^* = 5 \ln(T) - \mathcal{A}^*$;
  (3) define $f(\alpha |a,b) = \frac{\Gamma(a+b)}{\Gamma(a)\Gamma(b)} \left( \frac{\alpha - 1.01}{0.99} \right)^{\alpha -1} \left( \frac{2-\alpha}{0.99} \right)^{b-1}.$
   Accept $\alpha^{(g+1)}$ with probability $\min\left( \frac{\pi (\alpha^{(g+1)})}{\pi (\alpha^{(g)})} \times \frac{f(\alpha^{(g)}|\mathcal{A}^*,\mathcal{B}^*)}{f(\alpha^{(g+1)}|\mathcal{A},\mathcal{B})}, 1 \right)$.

Posterior for $\sigma$: $\sigma^{\frac{\alpha}{\alpha-1}} |{J^Y, U, \Theta} \sim \mathbb{IG} (\mathcal{A},\mathcal{B})$,
where $\mathcal{A} =  c^* +T$;
  $\mathcal{B} = C^* + \sum_{t=0}^{T-1} \left| \frac{J_{t+1}^Y}{\Delta^{\frac{1}{\alpha}} t_\alpha(U_{t+1})} \right|$;
  $c^*=2.5$ and $C^*=0.1$ are hyperparameters of the prior of $\sigma^{\frac{\alpha}{\alpha-1}}$.

  Posterior for $J^Y$ $\propto \pi (J_{t+1}^Y)$: $\pi (J_{t+1}^Y) := \exp \left\{ \frac{-J_{t+1}}{2(1-\rho^2) V_t \Delta} \left[ J_{t+1} - 2\left( H_{t+1} - \frac{\rho}{\sigma_v} D_{t+1} \right) \right]
  \right\} \times \exp \left\{ -\left| \frac{J_{t+1}}{\sigma \Delta^{\frac{1}{\alpha}} t_\alpha(U_{t+1})} \right|^{\frac{\alpha}{\alpha-1}} \right\} |J_{t+1}|^{\frac{1}{\alpha-1}}$;
 $D_{t+1} = V_{t+1} - V_t - \kappa (\theta - V_t)\Delta $;
  and $H_{t+1} = Y_{t+1} - Y_t - \left(r_t - \frac{1}{2} V_t + \Phi_J(-i) + \eta_s V_t\right)\Delta $.

Posterior for $U$ $\propto \pi(U_{t+1}) := f(U_{t+1}) \times \left[ \mathbf{1}_{J_{t+1} \in (-\infty,0) \cap U_{t+1} \in \left(-\frac{1}{2}, \frac{\alpha - 2}{2\alpha} \right)}  + \mathbf{1}_{J_{t+1} \in (0,\infty) \cap U_{t+1} \in \left( \frac{\alpha - 2}{2\alpha},\frac{1}{2} \right)} \right]$,
where $f(U_{t+1}) = \exp\left\{ -\left| \frac{J_{t+1}}{\sigma \Delta^\frac{1}{\alpha} t_\alpha (U_{t+1})} \right|^{\frac{\alpha}{\alpha-1}} +1 \right\} \left| \frac{J_{t+1}}{\sigma \Delta^\frac{1}{\alpha} t_\alpha (U_{t+1})} \right|^{\frac{\alpha}{\alpha-1}}$. We update $U_{t+1}^{(g+1)}$ with the following steps:
   (1) if $J_{t+1}<0$, draw $U_{t+1}^{(g+1)}$ from $\mathbb{U}(-\frac{1}{2}, \frac{\alpha - 2}{2\alpha})$; if $J_{t+1}>0$, draw $U_{t+1}^{(g+1)}$ from $\mathbb{U}( \frac{\alpha - 2}{2\alpha}, \frac{1}{2})$;
  (2) draw $u$ from $\mathbb{U}(0,1)$;
  (3) accept $U_{t+1}^{(g+1)}$ if $u<f\left( U_{t+1}^{(g+1)} \right)$.

\end{appendices}

\end{document}